# Rings and Halos in the Mid-Infrared: The Planetary Nebulae NGC 7354 and NGC 3242

J.P. Phillips[1], G. Ramos-Larios[1], K.-P. Schröder[2], & J. L. Verbena Contreras[2]

[1] Instituto de Astronomía y Meteorología, Av. Vallarta No. 2602, Col. Arcos Vallarta, C.P. 44130 Guadalajara, Jalisco, México   e-mails : jpp@astro.iam.udg.mx, gerardo@astro.iam.udg.mx

[2] Departamento de Astronomía, Div. CNyE, Universidad de Guanajuato, A.P. 144, Guanajuato, C.P. 36000, México, e-mails: kps@astro.ugto.mx, jluis@astro.ugto.mx

**Abstract**

We present images of the planetary nebulae (PNe) NGC 7354 and NGC 3242 in four mid-infrared (MIR) photometric bands centred at 3.6, 4.5, 5.8 and 8.0 µm; the results of observations undertaken using the Spitzer Space Telescope (SST). The resulting images show the presence of a halo and rings in NGC 3242, as previously observed through narrow band imaging at visual wavelengths, as well as evidence for a comparable halo and ring system in NGC 7354. This is the first time that a halo and rings have been observed in the latter source. Similarly, whilst a previous detection of partial rings has been suggested for NGC 3132 (Hora et al. 2004), the present observations appear to constitute the first detections of complete ring systems outside of the visual wavelength regime. The halo/core emission ratios appear to be preferentially higher at MIR wavelengths than is the case in the visible, and show less steep fall-offs than those observed in the [OIII] $\lambda\lambda$ 4959+5007 Å transitions. The variation in surface brightness S with radial distance R, where this is approximated by the power law relation $S \propto R^{-\beta}$, implies exponents $1.7 < \beta < 4$ in the inner portions of the halos, and $4.5 < \beta < 12$ towards the outer limits of these structures. The value of $\beta$ is also, in most cases, somewhat smaller at longer wavelengths.

It is additionally noted that the 3.6µm/4.5µm, 5.8µm/4.5µm and 8.0µm/4.5µm flux ratios increase markedly away from the nuclei, and reach their maximum values in the halos, where much of the longer wave flux may derive from dust band and continuum emission. An analysis of the rings suggests that some of them, at least, are likely to be associated with higher densities of dust particles, and that the gaseous and particle ring systems are likely to be spatially in register.

Finally, we have analysed the formation of halos as a result of radiatively accelerated mass loss in the AGB progenitors. Although the models assume that dust formation occurs in C-rich environments, we note that qualitatively similar results would be expected for O-rich progenitors as well. The model fall-offs in halo density are found to result in gradients in halo surface brightness which are similar to those observed in the visible and MIR.

**Key Words**: (ISM:) planetary nebulae: individual: NGC 3242, NGC 7534 --- ISM: jets and outflows --- infrared: ISM --- stars: AGB and post-AGB

# 1. Introduction

The envelopes of planetary nebulae (PNe) undergo various marked changes during their brief evolutionary lifetimes. Thus for instance, as central stars evolve to higher temperatures, and the fluxes of ionising photons increase, then D-type ionisation fronts are expected to advance into the inner portions of the AGB envelopes. This leads to appreciable changes in the pressure and density characteristics of these regimes, and wipes out much of the evidence relating to later phases of AGB mass loss (e.g. Meijerink et al. 2003; Perinotto et al. 2004; Schönberner et al. 2005, Mellema 1994).

Subsequent interaction of the envelopes with fast radiatively driven stellar winds then leads to high temperature bubbles of shocked stellar wind, together with denser, lower temperature regimes of shocked AGB plasma; the latter constrained by a contact discontinuity at smaller radii, and a forward leading shock (e.g. Schönberner et al. 2005; Schönberner & Steffen 2002). Evidence for these bubbles has become apparent from X-ray observations by the Chandra observatory, whence it is clear that temperatures are of order several millions of degrees (e.g. Yu et al. 2009; Kastner et al. 2008).

As the shells expand further, leading to lower mean plasma densities, and the central star evolves to even higher effective temperatures, then ionisation of the remaining AGB envelope occurs over time scales of $\sim$ few x$10^2$ years (Steffen & Schönberner 2003), and leads to the triple-shelled configuration noted in many more evolved PNe. The nebulae contain an inner bright rim, the fruits of interaction with the stellar wind; an inner shell surrounding this rim, the artefact of earlier ionisation by D-type fronts; and a more extended halo representing the original AGB mass-loss envelope (see e.g. Fig. 1, where the rim, shell and halo of NGC 3242 are explicitly labelled). However, this is by no means the final word in the evolution of the shells.

Eventually, a critical phase is reached at which central star hydrogen burning stops, stellar luminosities and temperatures decline, and there occurs a precipitous reduction in the fluxes of ionising photons. This leads to a contraction of the Stromgren regime within which ionization occurs, and the development of recombination halos, such as have been previously described by Phillips (2000), Tylenda (1986) and

Corradi et al. (2000). Such recombination halos can easily be confused with the more normal AGB halos described above, although they normally occur where shell surface brightnesses are lower, levels of nebular excitation are reduced, Zanstra temperatures are small and $T_Z(HI) \cong T_Z(HeII)$. They also lead to "halos" which are brighter than the more common-or-garden AGB halos described above.

Subsequent evolution of the shells eventually leads to very low densities, comparable to the densities of the interstellar medium (ISM), and this may lead to distortion and disintegration of the envelopes through shearing and hydrodynamic instabilities (see our further comments below).

The AGB halos have relatively low densities compared to the inner shells and nebular rims, and this may imply that they are in non-thermodynmic equilibrium, and have high electron temperatures (e.g. Marten 1993; Monreal-Ibero et al. 2005). It has also been noted that whilst the inner shells have surface brightness of order ~0.1-0.01 those of the inner nebular rim, the brightnessess of the outer halos are an order of magnitude less, of order $< 2 \, 10^{-3}$ peak nebular values. Most of the halos are found to be circularly symmetric, or very slightly elliptical (e.g. Hsia, Li & Ip 2008). Exceptions to this occur where the envelopes interact with the interstellar medium, however, a process which results in distortion of the shells, the formation of bright low-excitation rims at the leading edges of the halos, and displacement of the central stars from the geometrical centres of the envelopes (see e.g. Tweedy & Kwitter (1994, 1996) for observational results, and Villaver et al. (2000), Soker & Zucker (1997), Dgani & Soker (1998) and Xilouris et al. (1996) for theoretical modelling of such interactions).

In the absence of such effects, however, it has been argued that the density and kinematic properties of the halos provide information on the mass-loss history of the progenitors (e.g. Steffen & Schönberner 2003); a situation which is all but impossible with the inner shell. This has led to various studies of "stellar wind paleontology", and the mass-loss history of the progenitors (e.g. Balick et al. 1992).

In addition to the theoretical and observational work which has gone into defining the global properties of these halos, it has also been noted that several such halos possess annular enhancements in

intensity – features which are commonly referred to as rings, and will be so described below.

Such rings have been detected as reflection artefacts in the neutral shells of young and proto- PNe, where they appear to arise from spherically symmetric enhancements in the density of outflowing dust (viz. the cases of CRL 2688 (Sahai et al. 1998), IRAS 17441-2411 (Su et al. 1998), IRAS 17150-3224 (Kwok et al. 1998), IRAS 16594-4656, IRAS 20028+3910 (Hrivnak et al. 2001)). Such regions are also present in the ionised halos of more evolved PNe as well, and have been detected in at least 8 nebular outflows, and a possible further 4 sources as well (viz. Terzian & Hajian 2000; Corradi et al. 2004). The nature of these rings is rather poorly defined, although it is clear that they are relatively faint, and have surface brightnesses which are no more than $\sim$ 15% of those of the underlying halos (e.g. Corradi et al. 2004; Hb 5 is an exception in that the ring-to-halo ratio is of order unity). The density enhancement of the rings compared to the inter-ring regime has also been found to vary from factors of $\sim$ 2 (Kwok et al. 2000) to as high as $\sim$ 10 (Mauron & Huggins 1999). Apart from this, our knowledge of the rings is relatively cursory, although at least one PNe (NGC 6543) has been investigated in greater depth (Balick et al. 2001; Hyung et al. 2001). The results for this source are not in all cases consistent, however. Thus, Balick et al. (2001) determine that the surface brightness properties of the rings can be modelled in terms of individual shells having negligible amounts of intervening gas. They also determine that electron temperatures within this regime are relatively normal, although line-widths are unusually broad (see also Bryce et al. 1992). Hyung et al. (2001), by contrast, find electron temperatures which are significantly larger; a phenomenon which may, if confirmed, arise from the formation processes responsible for the rings.

Despite the short observational history of this phenomenon, however, and the lack of evidence concerning the physical properties of the rings, there have nevertheless been several suggestions as to how such features might form.

Thus for instance, Simis, Icke & Dominik (2001) and Meijerink, Mellema & Simis (2003) have analysed viscous momentum coupling between outflowing gas and dust; an analysis in which dust grains are permitted to drift with respect to the neutral AGB envelope. The results show that spherically symmetric regions will develop in which dust-to-gas ratios are enhanced, and lead to results which are closely similar to those observed in young PNe.

Garcia-Segura, Lopez & Franco (2001), on the other hand, point out that where there are changes in the polarity of the progenitor magnetic field, and where the stellar wind is also magnetised, then this would lead to pressure oscillations which drive circularly symmetric compressions into the outflowing winds. Their results are startlingly similar to those observed for He 2-90, although it has been suggested that the magnetic dynamo cycle in AGB stars would be much shorter than that corresponding to the rings (Meijerink et al. 2003).

Mastrodemos & Morris (1999) point out that mass-loss within central binary systems has the potential to lead to spiral shocks, and, again, to ring-like density enhancements similar to those observed in many PNe, whilst such features may also derive from relaxation oscillations within the progenitor (Van Horn et al. 2003); solar-type cycles in magnetic activity (Soker 2000); and variations in the periods of variability in the AGB progenitors (Zijlstra, Bedding & Mattei 2002). Icke, Frank & Heske (1992) suggest a process in which the oscillatory pumping in the interior layers of an AGB star leads to multi-periodicity or chaotic motion in the surface layers, although it's not clear if this can really account for the observed periodicity of the rings (Soker 2000). Similarly, Harpaz, Rappaport & Soker (1997) have proposed that the periastron passage of a stellar companion might modulate progenitor mass-loss rates, although it seems that the exact circularity of the predicted rings may be inconsistent with observations (Sahai et al. 1998).

There is no overriding mechanism which can be claimed to be uniquely in accord with the properties of the rings – most of them, for instance, indicate similar ring/halo surface brightness ratios. It seems clear however that the periodicities and time-scale of the rings are inconsistent with mechanisms based on helium shell flashes, since the typical interflash period is > $10^4$ yrs (Sahai et al. 1998; Kwok et al. 1998).

An interesting consequence of certain of the models, such as those of Simis, Icke & Dominik (2001) or Mastrodemos & Morris (1999), is that a drift of grains with respect to the gas might be expected to lead to positional de-coupling of the dust and gaseous rings. Where the grains are responsible for warm dust continua, or, say, give rise to polycyclic aromatic hydrocarbon (PAH) emission bands, then infrared emitting rings might be displaced from their gasous counterparts (as observed through narrow band H$\alpha$, [NII] or [OIII] imaging). Alternatively, where 3.6 and 4.5 $\mu$m emission is associated with gaseous components of emission, as appears to be the case in many PNe (e.g. Hora et al. 2004; Phillips & Ramos-Larios 2008a, b; Ramos-Larios & Phillips 2008), but the 5.8 and 8.0 $\mu$m fluxes are dominated by emission from small PAH emitting grains, then yet again, one might see some evolution in ring placement and structure on passing from shorter to longer MIR wavelengths. We shall investigate such possibilities in our analysis below.

Previous MIR observations of partial ring structures have been reported for the case of NGC 3132 (Hora et al. 2004). We report here the further MIR observation of more complete ring systems in NGC 3242 and NGC 7354. The rings in NGC 3242 were first discovered in the [OIII] and H$\alpha$+[NII] imaging of Corradi et al. (2003, 2004). In the case of NGC 7354, by contrast, it was concluded that there was little evidence for a halo down to $\approx 2 \times 10^{-3}$ peak nebular intensity (Corradi et al. 2003). Our present results will show that this source does have a halo, and that the halo also contains rings.

It will be noted that the strength of the halo emission (when compared to peak core fluxes) is ~10 times greater than is observed for optical permitted and forbidden line transitions. This argues for markedly differing emission mechanisms in the MIR from those prevailing in the visible.

Finally, we shall note below that the fall-off of halo emission at 3.6 and 4.5 $\mu$m appears to consist of two main trends; an interior fall-off in surface brightness $S \propto r^{-\beta}$, where $\beta \sim 1.4 \rightarrow 4$, corresponding to the regime where the rings themselves appear to be located, followed by a much steeper decline in which $\beta \sim 6.3 \rightarrow 12$. The fall-off at longer wavelengths tends to qualitatively similar, although with values of $\beta$ which are somewhat less. The possible origins of these gradients are discussed in Sect. 4.

## 2. Observations

We shall be making use, in the following analysis, of data products deriving from SST program 30285 ("Spitzer observations of planetary nebulae 2"); the near infrared (NIR) 2MASS all sky survey; and HST program 8773 undertaken using the Wide Field Planetary Camera 2 (WFPC2).

The Spitzer observations of NGC 3242 and NGC 7354 took place on 29/12/2006 and 09/08/2006, after which the raw results went through various stages of analysis as described in the IRAC data handbook (available at http://ssc.spitzer.caltech.edu/irac/dh/iracdatahandbook3.0.pdf). The first of these processes results in the so-called Basic Calibrated Data (BCD). In this case, the raw observations are converted into an appropriately flux-calibrated image, and the primary instrumental defects are removed. A further stage of processing (post BCD) was then undertaken using a specific, conservative set of parameters, and included cosmetic corrections to the images – the removal of defects which are not based on well-established instrumental artefacts or detector physics. A so-called "pointing refinement" is also undertaken, whereby point sources within the fields are astrometrically matched to sources in the 2MASS catalogue, whilst mosaics are produced from the multiple AORs (Astronomical Observation Requests; in the case of Program 30285, this involved combining 14 BCDs for each of the nebulae investigated here).

The present results all correspond to post-BCD products and are, as a result, relatively free from artefacts; are well calibrated in units of MJy sr$^{-1}$; and have reasonably flat emission backgrounds. An exception to this are the weak central horizontal emission bands at 8.0 $\mu$m caused by the bright central cores. We have not removed these from the present results, since they have little impact upon the analysis or its interpretation. They do however need to be considered when obtaining profiles through the source.

The observations were taken using the Infrared Array Camera (IRAC; Fazio et al. 2004), and employed filters having isophotal wavelengths (and bandwidths $\Delta\lambda$) of 3.550 $\mu$m ($\Delta\lambda$ = 0.75 $\mu$m), 4.493 $\mu$m ($\Delta\lambda$ = 1.9015 $\mu$m), 5.731 $\mu$m ($\Delta\lambda$ = 1.425 $\mu$m) and 7.872 $\mu$m ($\Delta\lambda$ = 2.905 $\mu$m).

The normal spatial resolution for this instrument varies between ~1.7 and ~2 arcsec (Fazio et al. 2004), and is reasonably similar in all of the bands, although there is a stronger diffraction halo at 8 μm than in the other IRAC bands. This leads to differences between the point source functions (PSFs) at ~0.1 peak flux. The observations were obtained in August 2006 (NGC 7354) and January 2008 (NGC 3242), correspond to post-basic calibrated data, and have a spatial resolution of 1.2 arcsec/pixel.

We have used these data to produce colour coded combined images of the sources in the four IRAC bands, where 3.6 μm is represented as blue, 4.5 μm as green, 5.8 μm is orange, and 8.0 μm is indicated by red. Several of these results have also been processed using unsharp masking techniques, whereby a blurred or "unsharp" positive of the original image is combined with the negative. This leads to a significant enhancement of higher spatial frequency components, and an apparent "sharpening" of the image (see e.g. Levi 1974). Profiles through these sources have also been produced with the aim of evaluating the fall-off in surface brightness of the halo structures. This involved an initial correcting for the effects of background emission; a component which is present in all of the bands, but is particularly strong at 5.8 and 8.0 μm. The latter two bands are also prone to slight gradients in the background of the order of $5 \times 10^{-4}$ MJy/sr/pix (although actual gradients vary depending upon the source and waveband under consideration, and the direction of the slice). We have removed these trends by subtracting lineal ramps from the results – a procedure which is more than adequate given the limited sizes of the nebulae. Both of the sources also suffer from weak central emission bands at 8.0 μm, as described above (although this is barely visible in the case of NGC 7354). The strength of this contaminant is low (~0.4% of peak emission fluxes), however, and tends to be constant with x-axis displacement. We have chosen slices through the nebula which minimise the impact of this feature, and the band is likely to have zero-to-negligible influence upon our present results.

The results were subsequently processed so as to indicate the variation of 3.6μm/4.5μm, 5.8μm/4.5μm and 8.0μm/4.5μm ratios with distance from the nucleus. The rationale behind this is based on the fact that many PNe possess strong PAH emission bands at 3.3, 6.2, 7.7 and 8.6 μm, located in the 3.6, 5.8 and 8.0 μm IRAC bands. Furthermore these features, and their associated plateau components, show evidence for increased strength outside of the nebular cores, in the halo regions of interest to our present analysis. Given that the 4.5 μm band is usually dominated by bremsstrahlung continua and a variety of molecular and ionic transitions, it then follows that the variation of these ratios gives some insight into the importance of PAH emitting grains (see e.g. Phillips & Ramos-Larios 2008a, 2008b).

We have also obtained contour mapping of the 8.0μm/4.5μm and 5.8μm/4.5μm and 3.6μm/4.5μm flux ratios over the sources. This was undertaken by estimating the levels of background emission, removing these from the 3.6, 4.5, 5.8, and 8.0 μm images, and subsequently setting values at < $3\sigma_{rms}$ noise levels to zero. The maps were then ratiod on a pixel-by-pixel basis, and the results contoured using standard IRAF programs. Contour levels are given through $R_n = A10^{(n-1)B}$, where the parameters (*A, B*) are provided in the captions to the figures.

Some care must be taken in interpreting the flux ratio results, however. The problems with large aperture photometry are described in the IRAC data handbook, and relate in part to scattering in an epoxy layer between the detector and multiplexer (Cohen et al. 2007). This leads to the need for flux corrections as described in Table 5.7 of the handbook; corrections which are of maximum order 0.944 at 3.6 μm, 0.937 at 4.5 μm, 0.772 at 5.8 μm and 0.737 at 8.0 μm. However, the precise value of this correction also depends on the underlying surface brightness distribution of the source, and for objects with size ~several arcminutes it is counselled to use corrections which are somewhat smaller. The handbook concludes that "this remains one of the largest outstanding calibration problems of IRAC".

We have, in the face of these problems, chosen to leave the flux ratio mapping unchanged. The maximum correction factors for the 8.0μm/4.5μm and 5.8μm/4.5μm ratios are likely to be > 0.8, but less than unity, and ignoring this correction has little effect upon our interpretation of the results.

Finally, we have produced emission and ratio profiles for ring features found within the halos of these sources. This involves taking slices across the rings, and removing smoother and underlying components of emission. We have also employed a "jitter" procedure first introduced by Corradi et al. (2004), in which a source image is displaced by a certain number of pixels in four orthogonal directions. The original image is then divided by the four displaced images, and the resulting ratio maps combined. This leads to enhancement of the ring structures, and effective removal of the underlying halo emission.

The 2MASS all-sky survey was undertaken between 1997 and 2001 using 1.3 m telescopes based at Mt Hopkins, Arizona, and at the CTIO in Chile. Each telescope was equipped with a three-channel camera, each consisting of a 256x256 array of HgCdTe detectors, and this permitted simultaneous observations of the sky in the J(1.25 $\mu$m), H(1.65 $\mu$m) and $K_S$(2.17 $\mu$m) photometric bands. Details of the data bases employed, and procedures used in the analysis of the data can be found in Skrutskie et al. (2006). The 2MASS results for NGC 3242 and NGC 7354 were acquired using the NASA Infrared Science Archive (IRSA), combined into three band colour imaging, and directly compared with results deriving from the Hubble Space Telescope (HST) and SST. We have in this case represented J band fluxes as blue, H band fluxes as green, and $K_S$ emission as red.

The Hubble Space Telescope (HST) was launched in April 1990, and consists of a f/24 2.4 m Ritchey-Chretien reflector. The observations of NGC 7354 were obtained on the 29[th] July 2001 using the WFPC2 (Holtzman et al. 1995), and were acquired as part of program 8773. The fits results were downloaded from the Hubble Legacy Archive at http://hla.stsci.edu/, and processed and combined to yield colour coded images of the source. In this case, we have selected exposures taken with filter F502N (central wavelength $\lambda_C$ = 5013 Å, bandwidth $\Delta\lambda$ = 47 Å) corresponding to emission from the $\lambda$5007 transition of [OIII], and for which the total exposure time was $\Delta t$ = 700 s; filter F555W, for which $\lambda_C$ = 5410 Å, $\Delta\lambda$ = 1605 Å, and $\Delta t$ = 260 s, and which is dominated by the nebular continuum; and filter F658N for which $\lambda_C$ = 6585 Å, $\Delta\lambda$ = 20 Å, and $\Delta t$ = 989 s, and which corresponds primarily to the $\lambda$6584 Å transition of [NII]. These three filters are represented in our combined image of the source as (respectively) blue, green and red, whilst the spatial resolution is 0.10 arcsec/pixel.

Finally, the F658N filter is known to leak a certain amount of flux from the adjoining H$\alpha$ transition. We note however that emission from this filter is concentrated in a patchy circum-nebular waist that bears little resemblance to the visual continuum, or 6 cm radio mapping of the source (see e.g. Fig. 1, and Hjellming, Bignell, & Balick 1978; the authors are unaware of any comparative H$\alpha$ imaging). It is therefore unlikely that contamination by H$\alpha$ is appreciable, or will seriously affect the apparent distribution of [NII].

## 3. Observational Results

### 3.1 The Case of NGC 7354

#### 3.1(i) The Characteristics of Nuclear Emission

Visual observations of the bright central nucleus of NGC 7354 reveal it to have an ellipsoidal morphology, and to be composed of an inner rim with aspect ratio ~1.6 and major axis length $\cong$ 30 arcsec, and a fainter shell with major axis dimensions $\cong$ 33 arcsec and a morphology which is almost circular (aspect ratio ~ 1.1). These two components are probably best interpreted in terms of the bubble blown rim/D-type shell structures described in Section 1. The velocity pattern of the rim is consistent with what would be expected for a spheroidal shell, and implies inclination corrected expansion velocities of 24.5 km s$^{-1}$ in [OIII], and 27.0 km s$^{-1}$ for [NII] (Sabbadin, Bianchini & Hamzaoglu 1983).

The Zanstra temperatures summarised by Phillips (2003a) imply mean values of $T_Z$(HI) $\cong$ 52.5 kK and $T_Z$(HeII) $\cong$ 98.7 kK. Such disparities in temperature are usually taken to imply that the shell is optically thin to H ionising radiation, whilst the large value of $T_Z$(HeII) suggests a comparably large stellar effective temperature.

HST imaging of the source is illustrated in Fig. 1, deriving from a combination of HST exposures deriving from the Hubble Legacy Archive. We have combined results taken in [NII] (coloured red), continuum (green) and [OIII] (blue) (see Sect. 2 for details). The corresponding 2MASS and Spitzer IRAC results, by contrast, are illustrated in the upper central and right-hand panels of Fig. 1, and again correspond to the combined results for all of the photometric bands, as described in Sect. 2. The latter results have also been processed using unsharp imaging techniques, which tends to emphasise finer and fainter aspects of the nebular structure.

The various [NII] features in the inner shell of NGC 7354, evident in the HST imaging in Fig. 1, are identifiable with the Fast Low-Excitation Regions (FLIERS) analysed in a series of papers by Balick and colleagues (Balick et al. 1993, 1994, 1998; Hajian et al. 1997); structures whose origins remain in doubt, but which have been observed in some ten or so PNe to date (a couple of these are labelled in Fig. 1). Whilst the kinematics of the NGC 7354 FLIERS are poorly determined, Hajian et al. (1997) consider that the unusual line emission properties of these structures are more likely to arise from photoionisation rather than shocks. Whatever their origins, however, it would seem that at least certain of the FLIERS may be enhanced in the infrared as well, including features in the upper right-hand corner of the interior shell, apparently visible in the NIR (2MASS) and Spitzer (MIR) results. These are located at ~15 arcsec from the central star, and along a PA of 310°. It is also possible that the jet-like extensions, at the major axis limits of the inner rim, may be visible in the Spitzer image of the source. Having said this, however, it should be noted that the relatively poorer resolutions of the 2MASS and Spitzer results may imply that at least certain of these features correspond to unrelated background sources.

A further characteristic of interest in the Spitzer imaging of this source is the red colouration of the halo, outside of the bright (and more elliptical) inner rim. This arises because of enhanced 5.8 and 8.0 $\mu$m emission, as discussed in our more detailed analysis below, and may indicate the presence of small PAH emitting dust particles outside of denser portions of the core.

## 3.1(ii) Halo Emission in NGC 7354

A larger scaled image of the source is illustrated in Fig. 2, where we have again combined colour coded IRAC band results, as described above, and processed the results using unsharp masking techniques. It will be seen that the bright central shell is surrounded by a much fainter and circular halo. This halo has a diameter of ~110 arcsec, contains at least three inner ring structures, and may also be surrounded by a much larger and more diffuse shell with diameter ≈250 arcsec. The latter structure is seen very faintly and partially at 5.8 μm, but is at its clearest (and strongest) at 8.0 μm.

This outermost shell may represent unrelated line of sight emission, or components of the ISM which are being ionised by the PN central star. We would argue, however, that neither of these is likely to be the case. Not only is the emission centred on NGC 7354, and appears roughly circular in appearance, but it also appears to be separated from the interior halo by a circular and lower intensity regime. It may therefore be that this outer shell was emitted during the earliest phases of AGB mass-loss; an event which was followed by a relative lull, and the final superwind phase of evolution.

Profiles through the source are illustrated in Fig. 3, where the several components of envelope emission are again in plain evidence. In this example, and for all of the other profiles to be considered here, the zero point corresponds to the central star position. The directions and widths of the slices are indicated in the figure captions, as are the positions corresponding to negative axial displacement (or relative position – hereafter referred to as RP). The surface brightnesses correspond to an average of pixel values between the limits of the slice, and over a direction which is orthogonal to that of the slice.

Emission in the high intensity central emission plateau, corresponding to the region of the wind-blown rim, falls-off steeply at its edges over a distance of ~10 arcsec. This region of surface brightness fall-off arises in the surrounding, more circular shell noted in Fig. 1, and corresponds to material which may have been ionised by a D-type front (see Sect. 1). There follows, after this, a marked moderation in the fall-off of shell surface brightness, in a region corresponding to the inner portions of the outer halo. The radial exponent β in this region has a value close to ~ 0. The surface brightness then again falls-off steeply in all of the wavebands, as is further illustrated in Fig. 4.

We have, for the latter figure, represented radial trends in surface brightness for the three longest wavelength MIR channels, and for three radial slices starting at the nucleus. The scatter in the results is an indication of the variations in fall-off for the differing radial directions, rather than representing photometric uncertainties in the measured fluxes.

It can be seen from this that although there is a rapid steepening in gradients with distance from the nucleus, one can identify three primary gradients in the decline of the halo surface brightness. At 4.5 μm for instance (and the results are also very similar at 3.6 μm as well), the exponent β varies from ~0 for relative positions (RPs) < 32 arcsec (i.e. log (RP/arcsec) < 1.51), to ~ 4 for 32 < RP < 50 arcsec, and β ~ 12 for RP > 50 arcsec. This last value is certainly very steep, and it could be argued that a simple cut-off in the halo at RP = 50 arcsec, allied to the instrumental response function, might lead to somewhat similar declines in intensity. However, our analysis of point source functions (PSFs) associated with stars within the field suggests that values of β would, in that case, approach something closer to ≈ 20. We would therefore suggest that this final steep decline is likely to be real, and may testify to an initial rapid increase in dM/dt at the onset of progenitor superwind mass-loss.

By contrast, the surface brightnesses at 5.8 and 8.0 μm fall-off somewhat less steeply, with β varying from 0-to-3-to-10 within the same ranges of distance from the central star.

It is not entirely clear what emission mechanisms are at work within the halo, although some insight may be gleaned from ratioing the various photometric bands. These ratios are illustrated in Fig. 4, and in the upper three panels of Fig. 5. It is clear, from the former diagram, that nuclear levels of 3.6, 4.5 and 5.8 μm emission are rather similar, and lead to ratios 3.6μm/4.5μm and 5.8μm/4.5μm which are close to ~0.3→0.5.

All of these ratios change markedly as one exits from the nucleus, however. In the first place, it is clear that the 8.0μm/4.5μm and 5.8μm/4.5μm ratios increase rapidly as one passes through the inner bright shell, up to a distance of 24 arcsec from the nucleus. This also occurs, in a more modest way, for the 3.6μm/4.5μm ratio as well, although this latter variation is less apparent in Fig. 3. Finally, the rate of change in the ratios moderates greatly at distances > 24 arcsec from the nucleus, where they take the typical values 8.0μm/4.5μm ~ 6, 5.8μm/4.5μm ~ 1.8, and 3.6μm/4.5μm ~ 0.9.

This variation in ratios is seen even more clearly in the upper three panels of Fig. 5, where we illustrate ratio variations over the projected surface of the source. One may note here the low ratios associated with the inner rim, the intermediate ratios connected with the inner shell, and the roughly uniform and higher ratios in the outer halo.

Apart from this, we note that this halo has not been observed in Hα+[NII] down to very low levels – to surface brightnesses which are < 2 $10^{-3}$ of the central source intensity. Corradi et al. (2003), on the basis of this, concluded that the halo in this source likely didn't exist at all. Exist it does, however, as our present results testify, and at an MIR surface brightness which is, at its brightest, ~0.07 of that in the central core.

So it is apparent that flux ratios are varying markedly with position in the source, and that relative levels of fluxes in the MIR are quite different from those in the visible – the halos are relatively much more bright. The same pattern of behaviour applies for NGC 3242 as well (see below), and we shall discuss some of the mechanisms which might be responsible for this in Sect. 3.2(ii).

**3.1(iii) The Ring System in NGC 7354**

Finally, we have taken a radial slice through the SE sector of the halo, and fitted the underlying emission with a sixth order least squares polynomial fit. This smoother, underlying fall-off is then removed from the total halo emission, leaving the peaks corresponding to the ring components alone. The results are illustrated in Fig. 6 for the four MIR photometric channels. Note here that although the median fall-off in halo surface brightness can be well approximated by a series of power-law trends (see Sect. 3.1(ii)), it is clear that this fall-off is subject to small-scale fluctuations which derive from intrinsic variations in the structure of the region. This is responsible for much of the scatter in Fig. 4. It is therefore not particularly useful to apply such laws to remove the underlying (and smoother) components of emission – particularly given that the rings themselves are relatively weak.

An aspect of this analysis which is of particular interest arises from the possibility that particle drift could lead to positional decoupling of the dust and gaseous rings (see Sect. 1). Where the emission mechanisms of the 3.6 and 8.0 $\mu$m ring components differ, therefore, such that the 3.6 $\mu$m flux is dominated by plasma and line components of emission, and that at 8.0 $\mu$m arises from small PAH emitting grains, then one might see some relative displacement in the rings in these differing wavelength regimes.

This may, indeed, be what is occurring in the present source – although if this is the case, then the degree of decoupling must be small. The short and longer wavelength rings are in approximate register, and any separation of the peaks is likely to be < 0.25 times the inter-ring spacing. This will also be found to be the case for NGC 3242 (see Sect. 3.2(iii)).

Such a result, should it be confirmed, may imply that the dust pattern is not drifting to any appreciable degree compared to its gaseous counterpart - or that both the shorter and longer wave components of MIR flux are dominated by dust continuum and/or band emission (see our further comments in Sect. 3.1(ii)), and the shorter and longer wave ring systems arise from the same annular patterns of particle enhancement.

Finally, we note that the strength of the 8.0 and 5.8 $\mu$m rings, compared to those at shorter wavelengths, suggests that the rings have larger levels of emission at longer MIR wavelengths. Where the longer wave bands are dominated by PAH band emission, then this presumably implies higher concentrations of the PAH emitting grains.

We have also represented the fractional emission in the rings in the lower panel of Fig. 6 – that is, the intensity I(R) in the rings divided by the combined emission I(R+H) in the ring and underlying halo. It can be seen from this that all of the rings have fractional peaks which are comparable to what is observed in the visible (<I(R)/I(H)> ~ 0.15 where excludes Hb 5; see Corradi et al. 2004). The outermost rings have fractional 8 μm peaks which are significantly smaller than what is found at 4.5 and 3.6 μm, however. It follows that whilst these longer wave peaks have fluxes indicative of enhanced particle densities, they are less strongly defined than is the case at shorter MIR wavelengths. On the other hand, something quite different appears to be occurring within the innermost ring of the sequence, at RP ≅ 27.8 arcsec. In this case, not only is the absolute level of 8.0 μm emission very much greater (see top panel of Fig. 6), but the ratio I(R)/I(R+H) ~ 0.1 is close to what is observed in other wavebands as well.

## 3.2 The Case of NGC 3242

### 3.2(i) The Characteristics of Nuclear Emission

NGC 3242 is not exactly a carbon copy of NGC 7354, but it is the next best thing. Suffice to say that it has Zanstra temperatures $T_Z(HI) \cong 56.3$ kK and $T_Z(HeII) \cong 89.9$ kK which are closely similar (Phillips 2003a; see also the detailed analysis of the central star temperature given by Pottasch & Bernard-Salas (2008)); and the kinematic structures of the shell/rim are comparable, and imply somewhat similar velocities of expansion ~ 20-30 km s$^{-1}$ (Meaburn, Lopez & Noriega-Crespo 2000; Balick, Preston & Icke 1987). It also has comparable evidence for FLIERS/jets along the major axis of the inner shell, features which have been investigated by Balick et al. (1993) (one of these is labelled in Fig. 1).

The HST, 2MASS and Spitzer images of the source are illustrated in Fig. 1; where the HST image is adapted from a version created by Bruce Balick & colleagues, available at http://www.spacetelescope.org/images/html/opo9738c3.html. In this case, emission in the filters F658N ([NII]), F502N ([OIII]), and F469N (He II λ4686 Å) are indicated respectively as red, green and blue. We have also created the 2MASS and Spitzer results by processing original J, H, K$_S$, and IRAC MIR images as described in Sect. 2. Note again that the 2MASS and Spitzer images correspond to unsharp masked results, designed to reveal finer details of the nebular structures.

Several things may be noted from an inter-comparison of the images. Firstly, the inner bright ring is better resolved (or more sharp) than was the case for NGC 7354, and also appears to be more complete, but otherwise shows a similar morphology and aspect ratio. Second, the inner shell about the rim has a major axis dimension of ~40 arcsec, is elliptical, and is apparent in all of the wavebands, from the visual through to the MIR. It has a grey colour in the Spitzer and 2MASS imaging, indicative of reasonably strong emission in all of the bands.

### 3.2(ii) The Nature and Properties of Halo Emission

The outer halo of NGC 3242 is faintly visible in the HST results, but very much stronger in the MIR (Fig. 1, lower right-hand panel). One can see the presence in the latter image of the three interior rings, and evidence for disruption of the rings to the upper right and lower-left hand sides of the source – a disruption which is associated with the arcs or bubbles of emission noted in the HST image of the source.

The most interesting of the MIR images, however, is probably that shown in Fig. 2, where we see evidence for as many as four rings within the halo of the source. This compares with the five rings observed by Corradi et al. (2004) (albeit two of the latter identifications are only partial).

Finally, it will be noted that the outer left-hand limits of the halo show a fragmentary appearance which will be discussed, along with other evidence, in a later paper concerning structural deformities in halos (Ramos-Larios & Phillips 2009, in preparation).

Profiles and flux ratios through the nebula are illustrated in Fig. 7, and show very closely similar diagnostics to those which were noted for NGC 7354. There are however differences. Note for instance how the 3.6µm/4.5µm ratios are similar across the source, and of order ~ 0.6. They are not however constant, as noted in the contour map illustrated in Fig. 5, where it will be seen that the ratio increases gradually from the centre to the outer limits.

It is again apparent that the rim region has the lowest 3.6μm/4.5μm, 5.8μm/4.5μm, and 8.0μm/4.5μm flux ratios; the elliptical shell has intermediate values of these parameters; and that the largest ratios occur within the halo, where 5.8 μm and 8.0 μm emission is dominant. There is however, unlike the case of NGC 7354, some evidence for a global increase in flux ratios as one passes to larger radii within the halo.

Finally, we note that the inner portion of the halo has an intensity which is ≤ 0.15 that of peak values – again much larger than the ratio noted for Hα+[NII] and [OIII] (Corradi et al. 2003).

It is therefore pertinent to ask why the relative MIR halo intensities are so much greater than those in the visible – a situation which applies to both of the sources investigated here. And why do the 5.8μm/4.5μm and 8.0μm/4.5μm ratios in particular increase to larger distances from the nebular centre?

Some indication of what might be happening may be determined from a Spitzer spectrum of NGC 3242, taken as part of the SIRTF IRS Calibration Program. The observation was acquired on 16/12/2005 as part of program 1427, used a 3.6x57 arcsec$^2$ slit oriented at a position angle of 110.5°, and was centred at RA(2000) = 10h 24m 49.2s, Dec(2000) = -18° 32m 23s. This data has not been fully processed, and is not illustrated here. However, it is apparent that various MIR bands and continua are of importance for the central portions of the source. We note, in particular, the presence of a broad continuum within the 5.8 μm photometric band – a component which becomes of increasing importance in the 8.0 μm channel. This is probably the dominant contribution to longer wave MIR IRAC fluxes. The 8.0 μm channel also contains strong transitions of [ArII] and [ArIII] at 6.99 and 8.99 μm, and a possible dust band peaking at 7.48 μm, blended with [Na III] at λ7.319 μm. This band may be interpretable in terms of a combination of C-C stretching and C-H in-plane bending modes associated with PAH type particles; an interpretation which gains weight from the presence of a further (and much weaker) band at 5.94 μm, which may derive from corresponding C-O stretching modes (see e.g. Tielens (2005) for a discussion of the properties of these particles). On the other hand, we note that Smith & McLean (2008) appear not to have detected the 3.3 mm PAH band feature in this source (we are grateful to an anonymous referee for drawing our attention to this reference, and to the existence of the Spitzer spectrum described above). Although these results don't necessarily inform us of what may be happening in the halo, they do suggest that dust band and continuum emission may be important. Where this is the case, then it would also explain certain of the flux ratio variations noted above (and illustrated in Figs. 4 & 5).

The longer wave emission in this source appears to have lines peaking at 10.54, 25.2, and 25.8 μm, although there is little if any evidence for dust emission bands.

These results, should their interpretation be accepted, appear to fly in the face of previous abundance measurements for this source, implying as they do a C/O abundance ratio > 1. Pottasch & Bernard-Salas (2008), for instance, determine that C/O < 1 for the nuclear regions of NGC 3242; a result also agrees with most other independent determinations of this parameter (see e.g. the references cited in Pottasch & Bernard-Salas (2008), and the values cited by Phillips (2003b)). This outcome, together with arguments concerning N abundances, is used to suggest a progenitor mass < 1.5 $M_\odot$.

It therefore follows that this particular PN may be similar to sources in which emission characteristics are indicative of dual C- and O-rich chemistries, a problem which is still under active consideration (see e.g. Gutenkunst et al. 2008; Perea-Calderón et al. 2009). One possible reason for such joint compositions is that late thermal pulses at the end of the AGB may lead to changes in C/O ratios (Waters et al. 1998); although this is just one of six or so possible explanations for this phenomenon.

The reasons for the excess levels of halo emission at shorter wavelengths (3.6 and 4.5 μm) are less easy to understand. Why should halo/core surface brightness ratios be so high when compared to those observed in the visible? One possible reason may be that the visual images of NGC 3242 include both H$\alpha$ and [NII] $\lambda\lambda$ 6548 & 6583 transitions. It's clear that emission by [NII] may represent a very important contribution indeed – not least because of its prevalence in the FLIERS noted in the HST imaging in Fig. 1. This contribution might jump-up core/halo intensity ratios by factors of 2 or 3, compared to those expected for H$\alpha$ emission alone. It is also possible that there are particularly strong transitions within the 4.5 $\mu$m band, including lines such as Br$\alpha$, the forbidden lines [Ar VI] and [Mg IV] close to 4.53 $\mu$m, and the shock or fluorescently excited v=0-0 S(8) and S(9) lines of H$_2$ at $\lambda\lambda$4.69 and 5.05 $\mu$m. Similar H$_2$ transitions may occur in the 3.6 $\mu$m channel as well.

Various PNe also appear to have hot dust within their halos, likely arising from stochastic heating of very small grains (e.g. Borkowski et al. 1994; Ramos-Larios & Phillips 2005; Phillips & Ramos-Larios 2005, 2006, 2007). This leads to characteristic J-H and H-K$_S$ indices in the near infrared, and may be associated with the self-same grains as are responsible for the PAH emission bands. The extension of these grain continua to longer infrared wavelengths might very well enhance 3.6 and 4.5 $\mu$m emission, and explain the strong increase in halo/core ratios noted above.

Similar processes are probably operable in the halo of NGC 7354, although it is possible that the halo in this source is neutral – a reason for its lack of detection in H$\alpha$. Under these circumstances, it is possible that PAH emission bands are particularly important, as noted in previous analyses of such regimes (see e.g. Phillips & Ramos-Larios 2008a, b; Ramos-Larios & Phillips 2008; and Ramos-Larios, Phillips & Cuesta 2008).

Finally, the variation of surface brightness with distance is illustrated in Fig. 8. The tendencies are again similar to those which were noted in NGC 7354, although rather less extreme. The inner portions of the halo are characterised by exponents $\beta \sim 1.7$ at 3.6 and 4.5 $\mu$m – and it is precisely in this regime that the rings are located. This then steepens to $\beta \sim 6.3$ for RP > 40 arcsec. Again, and as for NGC 7354, gradients are shallower for the 5.8 and 8.0 $\mu$m results, with $\beta$ in the outer region of the halo taking a value $\sim 4.5$.

We note that gradients for the inner parts of the halo have also been determined in the $\lambda\lambda$ 4959+5007 Å transitions of [OIII], where the value for $\beta$ was found to be 4.5 (Monreal-Ibero et al. 2005). This is clearly much steeper than is found in the MIR, and suggests that we need to be careful in interpreting such fall-offs. This question will be further addressed in Sect. 4.

It is worth noting that the halo discussed above is also surrounded by a much larger region of emission which may, or may not be related to NGC 3242 itself (e.g. Deeming 1966; Bond 1981; Zanin & Weinberger 1997). Some of the arguments relating to this envelope have been rehearsed by Meaburn et al. (2000) and Rosado (1986), whilst more recent INT wide field imaging by Corradi, and Galex images in the ultraviolet and visible have been made available on the internet (see the sites http://www.ing.iac.es/PR/science/n3242.jpg and http://photojournal.jpl.nasa.gov/catalog/PIA11968). Two of us will be discussing this region in a later paper (Ramos-Larios & Phillips 2009, in preparation), where we will conclude that the emission is unrelated to the NGC 3242 mass-loss process, and likely represents nearby material which is being ionised by the nebular central star. We shall also suggest that it is directly interacting with the halo about this source.

Finally, and as a general comment, we wish to point out that the similarities between these sources, noted in Sect. 3.2(i), which include the presence of FLIERS, high Zanstra temperatures, rings, and comparable core emission morphologies, are also shared by many other sources of their ilk. These similarities have not, to our knowledge, been commented upon before.

Thus for instance, fully half of sources having well-defined ring systems also appear to possess FLIERS, suggesting that the two phenomena may be intimately related. Similarly, it is worth noting that NGC 7354 has a rather modest Galactic latitude (b = 2.3°), whilst NGC 3242 is located at significantly higher latitudes (b = 32°). In this respect at least, therefore, the two sources are somewhat different. Taken as a whole however, we find that both of these groups of nebulae are characterised by large mean values of |b|, of order 16.0 ± 3.8 for the ten FLIER sources of Balick et al. (1993, 1994, 1998) and Hajian et al. (1997), and |b| = 14.8 ± 4.0 for the 13 clear and probable ring sources of Corradi et al. (2004). Such high mean Galactic latitudes are usually

taken to imply low mean progenitor masses (e.g. Phillips 2001), although some care must be taken not to over-interpret the present results. We note for instance that the FLIERS and rings constitute rather finely chiselled features which may prove difficult to discern at larger distances from the Sun. If the sources *are* located at smaller distances, then this will tend to increase their mean Galactic latitudes. Similarly, it is worth pointing out that halos are rather delicate by-products of stellar mass-loss history, and will be more easily destroyed by the ISM at lower heights above the Galactic plane. It is therefore possible that a variety of biases, physical as well as observational, may be determining the high apparent latitudes of these particular samples of sources.

### 3.2(iii) The Rings in NGC 3242

We have removed underlying, smoother components of halo emission in NGC 3242 using the techniques described in Sect. 3.1(iii). The results are shown in Fig. 9. We were only capable of rescuing two of the rings from these rather noisy results, however, those located at RPs of -31.4 and -40 arcsec.

The results are fascinating, and show that 8 $\mu$m emission is strong in both of the rings – much stronger than is the case in other wavebands - and that ratios I(R)/I(R+H) are in all cases comparable. We conclude from this that the rings in NGC 3242 are likely to have enhanced dust components of emission – a situation which is almost certainly the case at longer wavelengths, where PAH emission is likely to be strong, and may very well also be the case shorter wavelengths as well, where the strength of the emission may indicate the presence warm thermal grain continua (see Sect. 3.1(ii)).

The position of the 8 $\mu$m peak at RP $\cong$ -40.7 arcsec, when compared to the equivalent shorter wavelength peaks, might, at a stretch, suggest some slight displacement between the rings. Not too much confidence should be placed on this single result alone, however. A rather clearer understanding of what may be happening may be gained from Fig. 10, where we have removed the underlying continuum using the procedures of Corradi et al. (2004). This involves producing four versions of an image shifted 3 pixels to the left and right, up and down. The original unshifted image is then ratiod separately with the four shifted images, and the resulting maps are summed together. It is these mean ratio maps which are illustrated in Fig. 10 together with comparable [OIII] results taken from Corradi et al. (2004).

We have superimposed the narrow white arcs of Corradi et al. (2004), used to define the positions of peak emission in the [OIII] rings, on all three sets of exposures in the visible and MIR. It will be seen that there is no clear displacement between the results – the MIR rings are closely similar in position to those noted in the visible, although the outer partial ring of Corradi et al. (2004) appears to be undetected in the MIR. Given that this is the case, this therefore suggests that the gaseous density enhancements, responsible the ring features in the visible, are closely co-spatial with the particle density enhancements which we suppose to be responsible for the MIR results.

## 4. The Radial Variation in Surface Brightness in NGC 3242 and NGC 7354

We have noted that the halos for NGC 7354 and NGC 3242 have qualitatively similar fall-offs in intensity, which can be characterised through the use of two or three values of the radial exponent $\beta$. The inner regions of the sources have values $1.7 < \beta < 3$, for instance, and similar values have also been noted for NGC 650 (Ueta 2006; Ramos-Larios, Phillips & Cuesta 2008) and in the visible (see above). These exponents $\beta$ subsequently increase to values which are very much larger, in the region of $4.5 < \beta < 10$. The former gradients are similar to what would be expected from hydrodynamic modelling of AGB mass-loss, and the resulting decrement in electron densities (see e.g. Sandin et al. 2008; Schönberner et al. 2005; Schönberner & Steffen 2002; Steffen & Schönberner 2003).

We have also noted that the value of $\beta$ differs between the shorter and longer wave results, and with respect to at least one measure of $\beta$ acquired in the $\lambda\lambda$ 4959+5007 Å transitions of [OIII]. Thus, Monreal-Ibero et al. (2005) find $\beta \cong 4.5$ for the variation of [OIII] intensity within the inner halo of NGC 3242, whilst the 3.6 and 4.5 $\mu$m fall-offs are better represented by $\beta \cong 1.7$.

So there arises the interesting question of what these results might imply in terms of the variation in halo physical properties. The [OIII] and H$\alpha$ + [NII] fall-offs measured by Corradi et al. (2004), Monreal-Ibero et al. (2005) and Schönberner et al. (2005), for instance, indicate that 3.3 < $\beta$ < 4.5 for a variety of PNe. These may be directly dependent upon the variation of $n_e^2$ through the nebular shells, appropriately integrated along the line of sight, and/or may be influenced by ionisation stratification in the outer reaches of the sources. It is interesting, in this context, to note that Sandin et al. (2008) determine 2 < $\alpha$ < 8 where $n_e \propto r^{-\alpha}$.

Similar problems occur for the present MIR results. The fall-off at longer wavelengths, for instance, may be influenced by variations in grain number densities, in the grain size distribution n(a), and in the decreasing levels of photon extinction at larger distances from the central star. The 3.6 and 4.5 $\mu$m results, by contrast, may depend upon the presence of a variety of forbidden and permitted line transitions (see e.g. Sect. 3.1(ii)), 3.3 $\mu$m PAH emission, warm grain continua, and thermal bremsstrahlung emission.

So the situation, taken all-in-all, appears to be rather unclear. We shall consider below how such variations in surface brightness compare with those expected from mass-loss modelling of a C-rich progenitor, and where one assumes that halo emission derives from either grain bands and/or continua, or an ionised halo.

The progenitor star is assumed to have representative mass of 2.25 $M_\odot$; a value which falls at the lower end of the range expected for carbon rich, solar metallicity AGB models (see e.g. Marigo 2008; Wachter et al. 2002). Where masses are somewhat less than this (say < 1.5 $M_\odot$), then third dredge-up is unlikely to occur, and carbon abundances remain relatively low. Where masses are > 4$\rightarrow$5 $M_\odot$, on the other hand, then hot bottom CNO burning causes C surface abundances to decline (e.g. Herwig & Blocker 2000). Detailed analyses of these and other mechanisms can be found in Busso et al. (1999) and Lattanzio & Wood (2004).

We have chosen the range of C-rich AGB progenitors because radiatively driven mass-loss appears to be more efficient in such stars:

the mean Planck pressure coefficients for amorphous carbon grains, for instance, are approximately five times as great as those for silicates, a disparity which requires significantly higher luminosities to produce O-rich stellar winds (Woitke 2006) - luminosities such as are found in massive hot-bottom burning O-rich AGB giants.

The naive assumption of a constant outflow with a constant mass-loss rate yields a density profile $r^{-2}$. Considering that the final, dust-driven ``superwind'' increases until shortly before its end (see, e.g., the models of Schröder et al. 1999), a steeper density profile is generally to be expected for the outer envelopes of PNe.

Hence, the question arises: can a plateau, followed by a much steeper density decline – such as suggested by the present observations of NGC 7354, with the case of NGC 3242 not being too much different - be explained naturally by the superwind-mass-loss history alone, or do we need more exotic physics to explain this?

In order to investigate this question, we here used mass-loss histories derived from a well calibrated and tested fast evolution code (Pols et al. 1997, 1998; Schröder et al. 1997), originally developed by Peter Eggleton at the IoA Cambridge (Eggleton 1971, 1972, 1973), which we have combined with the parameterized mass-loss rates of detailed models of dust-driven, C-rich winds by the Sedlmayr group in Berlin (see Schroeder et al. 1999 and references therein). These dynamical wind models, which required massive CPU-time on a Cray super-cumputer, include microscopic dust-formation chemistry (PAH-related), radiative transport (all time-dependent), and a periodic piston at the inner boundary to simulate the mechanical energy input for the extended giant chromosphere, below the dust-formation region.

From the time-averaged mass-loss rates of 50 models with different basic physical parameters of the supergiant (L, M, $T_{eff}$), a parameterized mass-loss description was then derived (Wachter et al. 2002).

This was employed by our evolution code at each time-step. The macroscopic, mutual feed-back between the changing physical properties of the supergiant and the mass-loss rate is what shapes the resulting mass-loss histories. A steep rise in the superwind mass-loss in the final 20-30,000 years results, as well as the right order of magnitude of the peak mass-loss rate arise as an unforced consequence of the analysis (see Fig. 11), without need of any ad-hoc parameters.

The dust-formation chemistry of O-rich winds (among the more massive TP-AGB supergiants) is much more complicated, and quantitative wind-models are still difficult to compute. But there is good reason to assume that O-rich chemistry leads to very similar mass-loss histories, since there is a similarly strong dependence of the mass-loss rate on the basic stellar parameters, and the comparable magnitude of the mass-loss ensures a similar influence on the physical properties of the TP-AGB supergiant.

To derive the resulting density profiles, we start with the simplest of possible assumptions:

(i) A constant outflow velocity.

(ii) No interaction with the surrounding ISM.

(iii) No alteration of the density profile by the central star radiation fields or winds.

Hence we ignore all possible, complex hydrodynamic processes and look only at the expansion of the dust-driven mass-loss by the simple equation of continuity. And there are good reasons for such a naive approach: the expansion velocity of a dust-driven wind does not vary quickly and vastly with time, and the outer parts of many PNe appear undisturbed by the fundamental changes in their centres. At the same time, their density is sufficiently larger than that of the surrounding ISM to suggest that any interaction with the latter only leads to secondary effects on the density profile.

A look at a typical mass-loss history (see Fig. 11) immediately suggests that a plateau-like density distribution could arise from the falling post-peak mass-loss, while the steeply increasing pre-peak superwind mass-loss should produce a very steep, outer density profile - just as observed for NGC 7354 (Fig. 4). We have assumed here a distance of 1.19 kpc based upon the statistical result of Phillips (2004) (a very similar distance is also obtained by Zhang (1995)), whence the innermost result, at 25" (1.4 dex) from the centre, corresponds to a physical radius of 4.46 $10^{17}$ cm (log $r_1$ = 17.65), the outermost (63" or 1.8 dex) to 1.12 $10^{18}$ cm (log $r_2$ = 18.05). We have also assumed an expansion velocity of 26 km s$^{-1}$ based on a mean of the [OIII] and [NII] results of Sabbadin, Bianchini, & Hamzaoglu (1983); a value which likely corresponds to a severe upper limit for the halo expansion itself. Given such a value, then the radial range in Fig. 12 would correspond to a mass-loss history of 8200 years. About 2/3 of this history is pre-peak, and 1/3 is post-peak. During this episode alone, our TP-AGB supergiant model (Fig. 11), with $M_i$ = 2.25 $M_\odot$, loses 0.4 $M_\odot$ to the halo wind. If we dilute that into the above radial range (see Fig. 12) then densities drop to o($10^{-22}$) g/cm$^3$.

Where 4.5 μm emission is predominantly produced by optically thin, collisionally excited recombination, rather than by scattered light or grain thermal emission (although see our previous comments in Sect. 3.1(ii)), then every volume element contributes a flux proportional to the local density squared, and the observed flux at any radial point is simply the line-of-sight (LOS) integral over $n^2$. This steepens the radial flux profile when compared to the radial density profile. Thus, where the model density fall-off is as noted in the upper left-hand panel of Fig. 12, then one obtains the LOS-$n^2$ profile illustrated in the upper right-hand panel of this figure. This exhibits a nearly plateau-like inner part and a steep ~$r^{-4}$ to $r^{-7}$ outer part, a result which bears a striking similarity with our observations in Fig. 4. The shell of NGC 3242 is a similar, but more compact, younger case: If we adopt a distance of 0.5 kpc, similar to the expansion distances of Terzian (1997) and Jacoby (1980) (see also Pottasch 1996 and Mellema 2004), then the inner observed point at 18" (1.25 dex) corresponds to a physical radius of 1.35 $10^{17}$ cm (log $r_1$ = 17.125), and the outer (70", 1.85 dex) to 5.3 $10^{17}$ cm (log $r_2$ = 17.725). Similarly, where one assumes an outflow velocity of 27 km s$^{-1}$, similar to the mean of [NII] and [OIII] velocities for the nucleus of the source (Balick, Preston & Icke 1987; Weinberger 1989) – values which are again likely to constitute upper limits for the halo expansion velocity - then the observed flux-profile can be understood in terms of the same mass-loss history as for NGC 7354, but for an earlier phase of expansion (see the lower left-hand panel of Fig. 12). The lower right-hand panel of Fig. 12 shows the resulting LOS-$n^2$ profile (again the radius is given in arcsecs), a result which is qualitatively similar to the observed profile for NGC 3242 (see Fig. 8). Although the correspondence is not precise – there is no "break", or rapid change in model fall-off gradients such as appear to be indicated in the observations, for instance – there is a significant change in β, from values ~ -1.4 at smaller radial distances (log(r/arcsec) < 1.55) to ~ -4.2 at larger distances. On the other hand, where one adopts a slower outflow velocity for the outer halos than the 26→27 km s$^{-1}$ used here, then the pattern-scale of the projected mass-loss history would be reduced, and we would see an even steeper decline in the PNe outer halos – a result which is closer to the observational results noted above.

With the encouraging similarities found above, we may conclude that the observed density profiles are reasonably consistent with the superwind mass-loss history, and that in a first approximation, no complicated hydrodynamical effects are needed to understand these trends.

The present LOS-$n^2$ results are valid for ionised nebular outflows, and may be appropriate for the 3.6 and 4.5 μm results depending upon what the predominant emission mechanisms might be. Given this situation, it would be interesting to marry the present analysis to hydrodynamic modeling of fully ionised halos. However, we note that where halo temperatures are uniform (although see our comments in Sect. 1), and ionisation of the exterior halos occurs over a time-scale of ≈$10^2$ yrs (see Sect. 1), then it is likely that the structures would be homologously related to their non-ionised counterparts: whilst expansion velocities and size scales would be larger, the nature of the density fall-offs would be comparable.

Where dust emission is dominant at 5.8 and 8.0 μm (although this has yet to be proven), then the models would imply an exponent β of order ~0 at smaller radii, increasing to -3.3 at larger radii – assuming that the number density of dust grains is proportional to the gas density; that the variation n(a) in the number density of dust grains with grain radius a is invariant with r; and where excitation processes are invariant with distance from the nucleus. This represents a somewhat unlikely suite of properties. Nevertheless, it is clear that the fall-off in emission would be somewhat less than that determined where line and continuum emission processes dominate the MIR fluxes.

## 5. Conclusions

We have presented imaging of the planetary nebulae NGC 7354 and NGC 3242 in four MIR photometric bands, acquired through program P30285 undertaken with the Spitzer Space Telescope.

NGC 7354 has been shown, for the first time, to have a halo of diameter ~110 arcsec (and possibly larger), whilst the halo in NGC 3242 has a comparable appearance to that detected in the visible. The level of MIR emission in the halos, however, compared to those deriving from the nebular cores, appears to be an order of magnitude greater than is determined from permitted and forbidden line transitions. The reasons why this should be the case at shorter IRAC wavelengths (3.6 & 4.5 $\mu$m) are far from being clear, although it is plain that a mix of differing MIR transitions are capable of helping, including possible contributions from $H_2$, various ionic and forbidden lines, and the 3.3 PAH band and associated plateau component. It has also been suggested that very small grains, such as those responsible for the PAH bands, may also lead to warm NIR continua in the halos of a variety of other PNe. The extension of these continua to the MIR may explain some of the increase in halo fluxes noted here.

Flux ratios between the 3.6, 5.8 and 8.0 $\mu$m bands on the one hand, and the 4.5 $\mu$m band on the other, show increases by factors of up to ~ 5.5 on passing from the nebular nuclei through to the halos. This sharp increase in ratios may be attributable to the increasing role of PAH band transitions, to the wings of amorphous silicate features peaking at 9.7 $\mu$m; to broader warm dust continua; and/or to $H_2$ molecular transitions.

The fall-off of surface brightness with radius is dependent upon the IRAC band which is being considered. It is generally less steep at 5.8 and 8.0 $\mu$m, and implies radial exponents $\beta$ ranging from 0 to ~3 in the inner parts of the halos, where various emission rings are located, to $\beta$ ~4.5→10 at larger radii. These values increase to 1.7→4 and ~6.3→12 in the 3.6 and 4.5 $\mu$m bands. It is clear, in at least one case (NGC 3242), that the fall-off in the inner halo is much less steep than has been determined in [OIII].

Although the mechanisms responsible for the MIR emission are somewhat uncertain, as noted above, it is nevertheless evident that MIR gradients in the inner parts of the halo are in accord with those determined in the visible.

We have finally investigated the formation of such halos by AGB progenitors in which dust formation occurs in carbon rich environments, although we also note that similar results are likely to apply where the progenitors are oxygen rich as well. The resulting fall-off in halo densities, as determined for a series of simplifying (but reasonable) assumptions, are used to determine integrated line-of-sight variations in $n_e^2$. The latter profiles are perhaps most comparable with the surface brightness gradients observed for visual permitted and forbidden line transitions, since the nature of the emission mechanisms in the MIR remain somewhat uncertain. Nevertheless, we find that the model results for NGC 3242 and NGC 7534 are comparable to those observed in the MIR, and more generally, appear similar to visual surface-brightness fall-offs in other PNe as well. The line-of-sight variations in density n, by contrast, are perhaps more relevant where emission derives from dust grains – grains such as may be important in determining fluxes within the 5.8 and 8.0 $\mu$m channels. For this case, the exponents $\beta$ appear less, as is observed to be the case in our present results.

Apart from the global features of halo emission noted above, we also note that the inner halo of NGC 7354 contains three relatively narrowly defined rings. This is the first time that such rings have been discovered outside of the visual wavelength regime. The mechanisms responsible for halo emission are by no means clear, as noted above, and it is possible that shorter MIR wavelengths are dominated by warm grain continua. It is therefore possible that we are witnessing evidence, in these rings, for radially symmetric enhancements in particle density. This uncertainty apart, however, we note that the innermost ring appears to be particularly enhanced at 8.0 $\mu$m, and this likely testifies to higher concentrations of outflowing grains.

By contrast, the rings in NGC 3242 have been previously observed by Corradi et al. (2004) in H$\alpha$ + [NII], although they again show evidence for enhanced dust emission in at least two of the structures.

Comparison of the positioning of these rings, both with respect to those observed at other MIR wavelengths, and (for NGC 3242) those observed in the visible, suggests that they are all closely co-spatial. It is therefore clear that there is little slippage of the ring patterns, and that enhancements in gas density are probably co-spatial with those in particle number densities.

**Acknowledgements**

We thank an anonymous referee for his careful review of this work, and several valuable suggestions. This research is based, in part, on observations made with the Spitzer Space Telescope, which is operated by the Jet Propulsion Laboratory, California Institute of Technology under a contract with NASA. Support for this work was provided by an award issued by JPL/Caltech. In addition to this, the work makes use of data products from the Two Micron All Sky Survey, which is a joint project of the University of Massachusetts and the Infrared Processing and Analysis Center/California Institute of Technology, funded by the National Aeronautics and Space Administration and the National Science Foundation. The 2MASS data was acquired using the NASA/ IPAC Infrared Science Archive, which is operated by the Jet Propulsion Laboratory, California Institute of Technology, under contract with the National Aeronautics and Space Administration. Finally, various of the results for NGC 3242 were based on observations made with the NASA/ESA Hubble Space Telescope, and obtained from the Hubble Legacy Archive, which is a collaboration between the Space Telescope Science Institute (STScI/NASA), the Space Telescope European Coordinating Facility (ST-ECF/ESA) and the Canadian Astronomy Data Centre (CADC/NRC/CSA). GRL acknowledges support from CONACyT (Mexico) grant 93172, whilst KPS acknowledges financial support from CONACyT (Mexico) under project grant number 80804 (CB-2008).

# References

Balick B., Alexander J., Hajian A.R., Terzian Y., Perinotto M., Patriarchi, P., 1998, AJ, 116, 360

Balick B., Gonzalez G., Frank, A., 1992, ApJ, 392, 582

Balick B., Perinotto M., Maccioni A., Terzian Y., Hajian A., 1994, ApJ, 424, 800

Balick B., Preston H.L., Icke V., 1987, AJ, 94, 1641

Balick B., Rugers M., Terzian Y., Chengalur J.N., 1993, ApJ, 411, 778

Balick B., Wilson J., Hajian A. R., 2001, AJ, 121, 354

Bond H.E., 1981, PASP, 93, 429

Borkowski K.J., Harrington J.P., Blair W., Bregman J.D., 1994, ApJ 435, 722

Busso M., Gallino R., Wasserburg G. J.,1999, ARA&A 37, 239

Casassus S., Roche P. F., Aitken D. K., Smith C. H., 2001, MNRAS, 320, 424

Cohen M., Parker Q.A., Green A. J., Murphy T., Miszalski B., Frew D.J., Meade M. R., Babler B., Indebetouw R., Whitney B. A., Watson C., Churchwell E. B., Watson D.F., 2007b, ApJ, 669, 343

Corradi R.L.M., Sánchez-Blázquez P., Mellema G., Giammanco C., Schwarz H. E., 2004, A&A, 417, 637

Corradi R.L.M., Schönberner D., Steffen M., Perinotto M., 2003, A&A, 354, 1071

Corradi R.L.M., Schönberner D., Steffen M., Perinotto M., 2003, MNRAS, 340, 417

Deeming T.J., 1966, ApJ, 146, 287

Dgani R., Soker N., 1998, ApJ, 495, 337

Eggleton P.P., 1971, MNRAS 151, 351

Eggleton P.P., 1972, MNRAS 156, 361

Eggleton P.P., 1973, MNRAS 163, 179

Fazio G., et al., 2004, ApJS, 154, 10

Garcia-Segura G., Lopez J.A., Franco J, 2001, ApJ, 560, 928

Gutenkunst S., Bernard-Salas J., Pottasch S. R., Sloan G. C., Houck J. R., 2008, ApJ. 680, 1206

Hajian A.R., Balick B., Terzian Y., Perinotto M., 1997, ApJ, 487, 304

Harpaz A., Rappaport S., Soker N., 1997, ApJ, 487, 809

Herwig F., Blocker T., 2000, in Noels, A., Magain, P., Caro, D., Jehin, E., Parmentier G., Thoul, A., eds, Proc. 35th Liege International Astrophysics Colloquium, The Galactic Halo : From Globular Cluster to Field Stars. Institut d'Astrophysique et de Geophysique, Liege, Belgium, p.59

Hjellming R.M., Bignell R.C., Balick B., 1978, Sky & Tel., 56, 199

Holtzman, J.A., Burrows, C.J., Casertano, S., Hester, J.J., Trauger, J.T., Watson A.M., Worthey G., 1995, PASP, 107, 156

Hora J.L., Latter W.B., Allen L.E., Marengo M., Deutsch L.K., Pipher, J.L., 2004, ApJS, 154, 296

Hrivnak B. J., Kwok S., Su K. Y. L., 2001, AJ, 121, 2775

Hsia C.H., Li J.Z., Ip W.-H., 2007, arXiv0712.2639H

Hyung S., Mellema G., Lee S.-J., Kim H., 2001, A&A, 378, 587

Icke V., Frank A., Heske A., 1992, A&A, 258, 341

Jacoby G. H., 198, 2000, ApJS, 42, 1

Kastner J.H., Gatley I., Merrill K.M., Probst R., Weintraub D.A., 1994, ApJ, 421, 600

Kastner J.H., Montez R., Jr., Balick B., De Marco, O., 2008, ApJ, 672, 957

Kwok S., Su K. Y. L., Hrivnak B. J. 1998, ApJ, 501, L117

Kwok S., Su K. Y. L., Stoesz J., 2001, in Szczerba, R, Górny, S.K., eds, Ap&SS Science Library Vol. 265, Post-AGB Objects as a Phase of Stellar Evolution. Kluwer Academic Publishers, Boston/Dordrecht/London, p. 115

Lattanzio J.C., Wood P.R., 2004, in Habing, H.J., Olofsson, H., eds, Asymptotic Giant Branch Stars. A&A Library, Springer, p. 23

Latter W.B., Hora J.L., in Habing, H.J., Lamers, H., eds, Proc. IAU Symp. 180, Planetary Nebulae. Kluwer, Dordrecht, Holland, p. 254

Levi L., 1974, Computer Graphics and Image Processing, 3, 163

Marigo P., 2008, Mem. S. A. It., 79, 403

Marten H., 1993, A&A, 277, L9

Mastrodemos N., Morris M., 1999, ApJ, 523, 357

Mauron N., Huggins P. J. 1999, A&A, 349, 203

Meaburn J., López J.A., Noriega-Crespo A., 2000, ApJS, 128, 321

Meijerink R., Mellema G., Simis Y., 2003, A&A, 405, 1075

Mellema G., 1994, A&A, 290, 915

Mellema G., 2004, A&A, 416, 623

Monreal-Ibero A., Roth M.M., Schönberner D., Steffen M., Bohm P., 2005, ApJ, 628, L139

Neufeld D.A., Yuan, Y., 2008, ApJ, 678, 974

Perea-Calderón J.V., García-Hernández D.A., García-Lario P., Szczerba R., Bobrowsky M., 2009, A&A, 495, 5

Perinotto M., Schönberner D., Steffen M., Calonaci C., 2004, A&A, 414, 993

Phillips, J.P., 2000, AJ, 119, 2332

Phillips J.P., 2001, PASP, 113, 839

Phillips J.P., 2003a, MNRAS, 344, 501

Phillips J.P., 2003b, MNRAS, 340, 883

Phillips J.P., 2004, MNRAS, 353, 589

Phillips J.P., Ramos-Larios G., 2005, MNRAS, 364, 849

Phillips J.P., Ramos-Larios G., 2006, MNRAS, 368, 1773

Phillips J.P., Ramos-Larios G., 2007, AJ, 133, 847

Phillips J.P., Ramos-Larios G., 2008a, MNRAS, 383, 1029

Phillips J.P., Ramos-Larios G., 2008b, MNRAS, 386, 995

Pols O.R., Tout C.A., Schröder K.-P., Eggleton P.P., Manners J., 1997, MNRAS, 289, 869

Pols O.R., Schröder K.-P., Hurley J.R., Tout C.A., Eggleton P.P., 1998, MNRAS, 298, 525

Pottasch S.R., 1996, A&A, 307, 561

Pottasch S.R., Bernard-Salas J., 2008, A&A, 490, 715

Ramos-Larios, G., Guerrero, M., Miranda, L.F., 2008, AJ, 135, 1441

Ramos-Larios G., Phillips J.P., 2005, MNRAS, 357, 732

Ramos-Larios G., Phillips J.P., 2008, MNRAS, 390, 1014

Ramos-Larios G., Phillips J.P., Cuesta L. 2008, MNRAS, 391, 52

Rosado M.,1986, RMA&A, 13, 49

Sabbadin F., Bianchini A., Hamzaoglu E., 1983, A&AS, 51, 119

Sahai R., Trauger J. T., Watson A. M., et al. 1998, ApJ, 493, 301

Sandin C., Schönberner D., Roth M. M., Steffen M., Böhm P., Monreal-Ibero, A., 2008, A&A, 486, 545

Schönberner D., Jacob, R., Steffen, M., Perinotto, M., Corradi, R.L., Acker, A., 2005, A&A, 431, 963

Schönberner D., Steffen M., 2002, RevMexAA (Serie de Conferencias), 12, 144

Schröder K.-P., Pols O.R., Eggleton P.P. 1997, MNRAS, 285, 696

Schröder K.-P., Winters J.M., Sedlmayr E., 1999, A&A, 349, 898

Simis Y.J.W., Icke V., Dominik C., 2001, A&A, 371, 205

Skrutskie M. F., et al., 2006, AJ, 131, 1163

Smith E.C.D., McLean I.S., 2008, ApJ, 676, 408

Soker N., 2000, ApJ, 540, 436

Soker N., Zucker D.B., 1997, MNRAS, 289, 665

Steffen M., Schönberner D., 2003, in Kwok S., Dopita M., Sutherland R., eds, Proc. IAU Symp. 209, Planetary Nebulae: Their Evolution and Role in the Universe. Astron. Soc. Pac., San Francisco, p. 439

Su K. Y. L., Volk K., Kwok S., Hrivnak B. J., 1998, ApJ, 508, 744

Terzian Y., 1997, in Habing, H.J., Lamers, H.J.G.L.M, eds, Proc. IAU Symp 180, Planetary nebulae. Kluwer Academic Publishers, Holland, p. 29.

Terzian Y., Hajian, A. R., 2000, in Kastner, J.H., Soker, N., Rappaport, S., eds, ASP Conf. Ser. Vol. 199, Asymmetrical Planetary Nebulae II: From Origins to Microstructures. Astron. Soc. Pac., San Francisco, p. 33

Tielens A.G.G.M., 2005, The Physics and Chemistry of the Interstellar Medium, C.U.P., Cambridge, England

Tweedy R.W., Kwitter K.B., 1994, AJ, 108, 188

Tweedy R.W., Kwitter K.B., 1996, ApJS, 107, 255

Tylenda R., 1986, 156, 217

Ueta T., 2006, ApJ, 650, 228

Van Horn H.M., Thomas J.H., Frank A., Blackman E.G., 2003, ApJ, 585, 983

Villaver E., Manchado A., García-Segura G., 2000, RMxAC, 9, 213

Wachter A., Schröder K.-P., Winters J.-M., Arndt T., Sedlmayr E., 2002, A&A 384, 452

Waters L.B.F.M., Beintema D.A., Zijlstra A.A., et al., 1998, A&A, 331, L61

Weinberger R., 1989, A&AS., 78, 301

Xilouris K. M., Papamastorakis J., Paleologou E., Terzian Y., 1996,
39

A&A, 310, 603

Yu Y. S., Nasdew R., Kastner J. H., Houck J., Behar E., Soker N., 2009, ApJ, 690, 440

Zanin C., Weinberger R., 1997, in Habing, H.J., Lamers, H.J.G.L.M, eds, Proc. IAU Symp 180, Planetary Nebulae. Kluwer Academic Publishers, Holland, p. 290

Zhang C.Y., 1995, ApJS, 98, 659

Zijlstra A.A., Bedding T.R., Mattei J.A., 2002, MNRAS, 334, 498

**Figure Captions**

**Figure 1**

Comparison of HST, 2MASS and Spitzer images of the central regions of NGC 7354 and NGC 3242, where the HST image for NGC 3242 is from http://www.spacetelescope.org/images/html/opo9738c3.html, and the other results have been processed by the authors using online 2MASS and HST data bases, and the present SST results. For the case of the HST image of NGC 3242, blue emission corresponds to HeII $\lambda$4686 Å, green represents the [OIII] $\lambda$5007 Å transition, and red is [NII] $\lambda$6583 Å, whilst for the HST image of NGC 7354 we have designated [NII] $\lambda$6583 Å emission by red, $\lambda$5410 Å continuum emission by green, and the [OIII] $\lambda$5007 Å transition as blue. The J, H and $K_S$ 2MASS results are represented as blue, green and red, whilst the 3.6, 4.5, 5.8 and 8.0 $\mu$m Spitzer results are denoted by blue, green, orange and red.

Note that all of the images are directly comparable with each other, having as they do identical scales, positioning and orientation. Similarly, the 2MASS and Spitzer images have been processed using unsharp masking techniques, leading to sharper representations of fainter nebular structures.

Particular points to note include the presence of FLIERS in both sources, delineated by the red [NII] components of emission in the HST images, and including jet-like features at the limits of the major axes. Certain of these have been labelled. A few of the FLIERS may also have been detected in the 2MASS and Spitzer images of the sources. It is clear that both sources also have similar morphologies at all wavelengths, consisting of an elliptical inner rim and enveloping shell. The outer halo of NGC 3242 is visible in the lower right-hand panel, where it and the rim and shell features are explicitly labelled.

**Figure 2**

Larger field MIR images of NGC 3242 and NGC 7354, where it can be seen that both sources show evidence for ring structures in their halos. There may also be evidence for an external and detached diffuse ring about NGC 7354. The colour coding of the IRAC fluxes is as described in Fig. 1, whilst it should also be noted that both of the images have been processed using unsharp masking techniques.

**Figure 3**

The variation of surface brightness through the centre of NGC 7354 (upper panel), where it will be noted that the vertical axis is in logarithmic units. The direction and width of the slice is indicated in the inserted image. The lower panel, by contrast, shows the variation of 8.0μm/4.5μm, 5.8μm/4.5μm and 3.6μm/4.5μm flux ratios close to the central region of the source. It will be noted that the 5.8μm/4.5μm and 8.0μm/4.5μm ratios show particularly strong increases towards the halo (i.e. for |RP| > 14 arcsec). The rings appear to be slightly better defined in the north-easterly and south-westerly potions of the halo; an effect which may arise as a result of weak contaminating bands at 8.0 μm, positioned horizontally across the centre of the image. The slit is oriented in an approximate N-S direction (see inserted panel), and has a width of 4.7 arcsec. Negative positional offsets correspond to the southern side of the central star (upper right-hand side of the inserted image).

**Figure 4**

The variation of halo surface brightness in NGC 7354 as a function of distance from the nebular centre, where we have illustrated results along three radial cuts. The scatter in the results is largely a result of slight asymmetries in the halo, whilst the various lines are eye-fitted to the results.

**Figure 5**

Flux ratio maps for NGC 3242 and NGC 7354. It will be noted that both sets of maps are closely similar, and show evidence for much lower ratios within the inner rim-like structures (delineated using darker shades of grey), somewhat higher values for the interior shells, located just outside of the rims, and finally, very much higher ratios for the halos (where grey levels are also at their lightest). Where flux ratios $R_n$ are represented through $R_n = A10^{(n-1)B}$ then the contour parameters [A,B] in NGC 3242 are given by [1.3, 0.1384] for 3.6μm/4.5μm, [0.4, 0.175] for 5.8μm/4.5μm, and [0.4, 0.0828] for 8.0μm/4.5μm. The corresponding parameters for NGC 7354 are given by [0.2, 0.0625] for 3.6μm/4.5μm, [0.35, 0.0756] for 5.8μm/4.5μm, and [1, 0.0625] for (8.0μm/4.5μm).

**Figure 6**

A slice through the ring system of NGC 7354, where underlying emission has been deleted using a sixth order least-squares polynomial fit. The peaks therefore represent the excess emission associated with the ring features alone. The top panel indicates this emission in terms of MJy/sr, whilst the lower panel represents the ratio S(R)/S(R+H); where S(R) is the surface brightness of the ring, and S(R+H) is the surface brightness of the rings and halo combined. The slit is oriented in an approximate N-S direction, and is located in the southern portion of the halo (see inserted panel). Its width is 10.2 arcsec.

**Figure 7**

As in Fig. 3, but for profiles through the centre of NGC 3242. The slit is oriented in an approximate E-W direction (see inserted panel), and has a width of 4.2 arcsec. Negative positional offsets correspond to the eastern side of the central star (lower left-hand side of the inserted image).

**Figure 8**

The variation of surface brightness with radial distance from the central star in NGC 3242. Details are otherwise as indicated in Fig. 4.

**Figure 9**

Emission for two of the rings in NGC 3242, where details are otherwise as stated in Fig. 6. The slit is oriented in an approximate E-W direction, and is located in the eastern portion of the halo (see inserted panel). Its width is 6.7 arcsec.

**Figure 10**

3.6 and 8.0 μm images of NGC 3242, processed so to emphasise the halo ring structures, together with comparable [OIII] images taken from Corradi et al. (2004). The top panels indicate the original processed images, where darker shades of grey indicate higher levels of emission, whilst the lower panels contain superimposed circles delineating the positions of maximum emission. The three sets of superimposed circles are identical, and reveal that all of the rings, those in the visible, 8.0 μm and 3.6 μm, appear to have identical distances from the central star. There is therefore little evidence for displacement between the patterns.

**Figure 11**

Final mass-loss history for an initial stellar mass of $M_i$ = 2.25 $M_\odot$, solar composition, and C-rich dust-formation chemistry. Marks indicate the decreasing, actual mass of the TP-AGB supergiant. Both short- and long-term variations of the mass-loss rate are caused by thermal pulses.

**Figure 12**

Model results for NGC 7354 and NGC 3242. The upper left-hand panel shows the density profile of an evolved PN envelope matching NGC 7354 (logarithmic radial scale). The sharp variation stems from the final thermal pulse past the mass-loss maximum in Fig. 12, while the much steeper, outer decline reflects the steeply increasing pre-peak mass-loss. The upper right-hand panel, by contrast, shows the corresponding line-of-sight integration over density-squared, using the density-profile shown in the upper left-hand panel, where the radius is now given in arcsec for an adopted distance of 1.19 kpc. Note the similarity to Fig. 4. The lower panels represent the equivalent variation for NGC 3242, except that the lower left hand results are for an earlier phase of the expansion (by 4000 yrs), and the distance is taken to be 0.5 kpc. Note the similarity to Fig. 9.

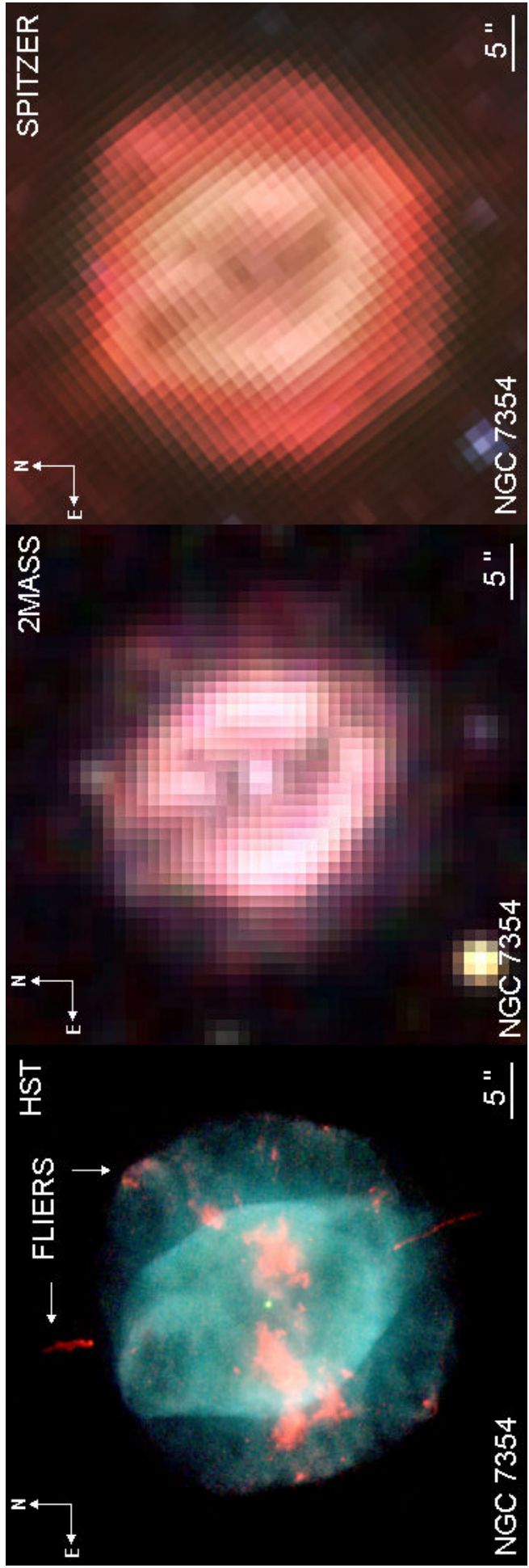
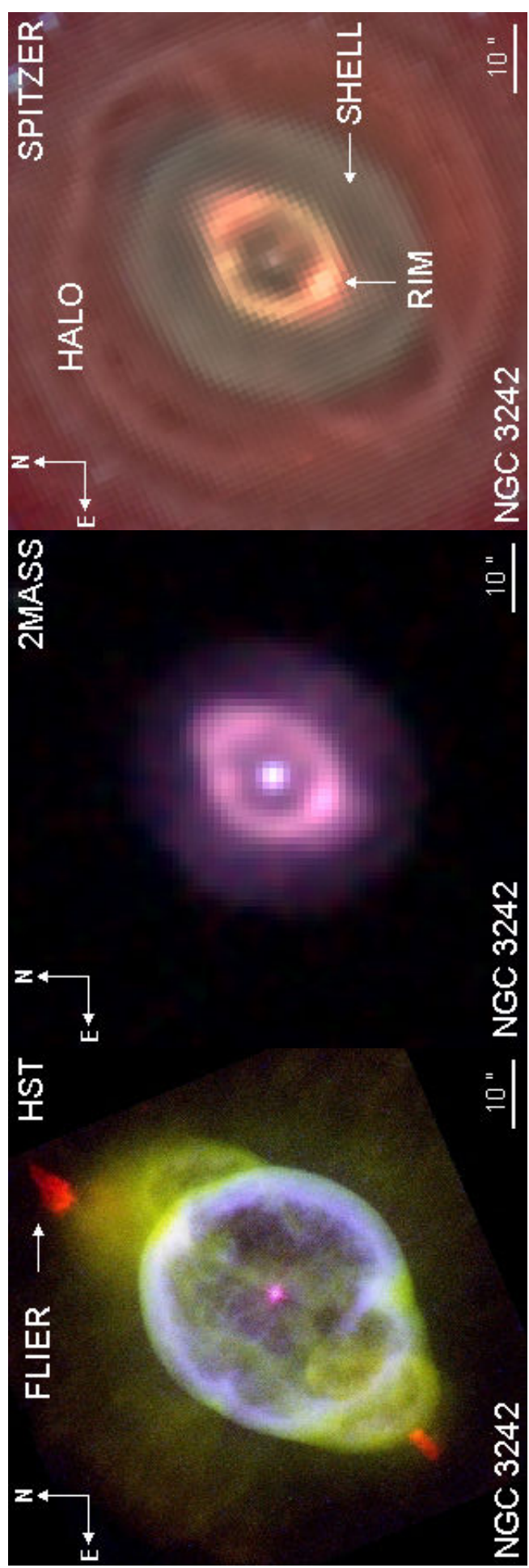

FIGURE 1

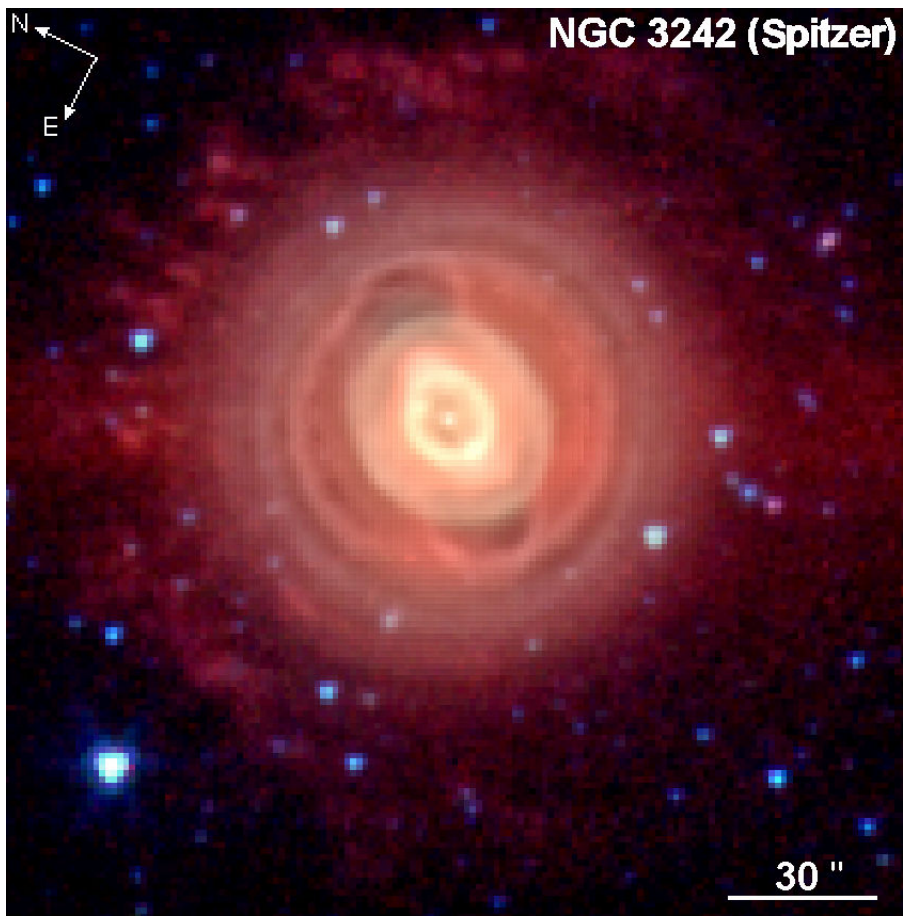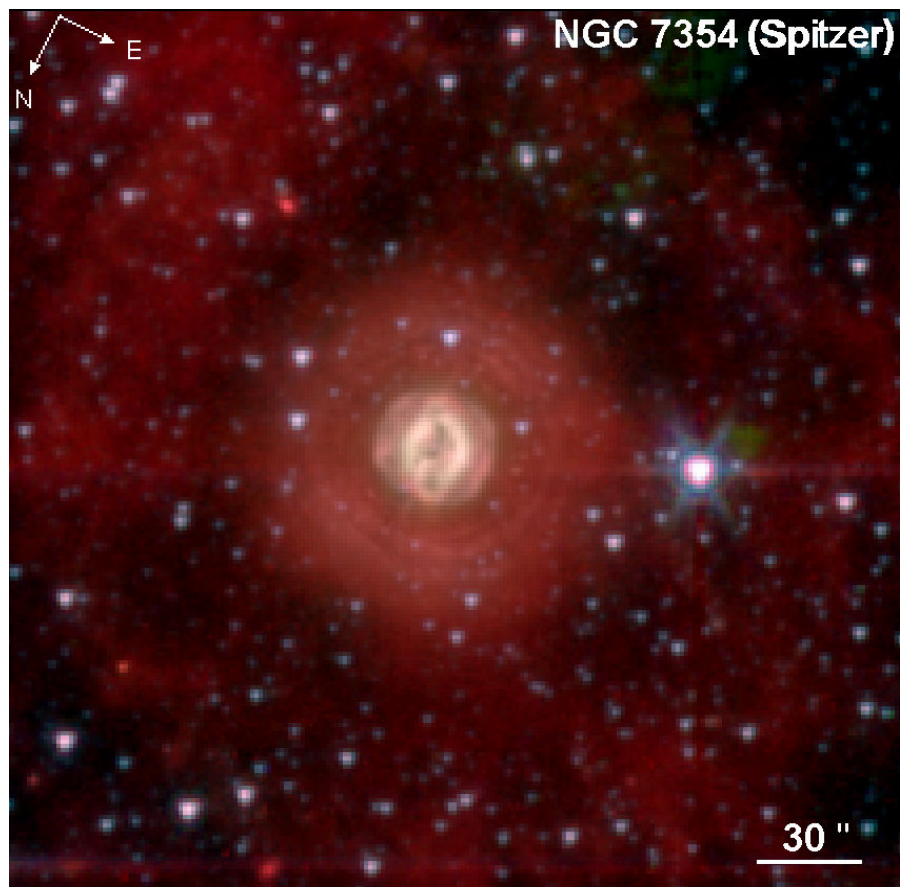

FIGURE 2

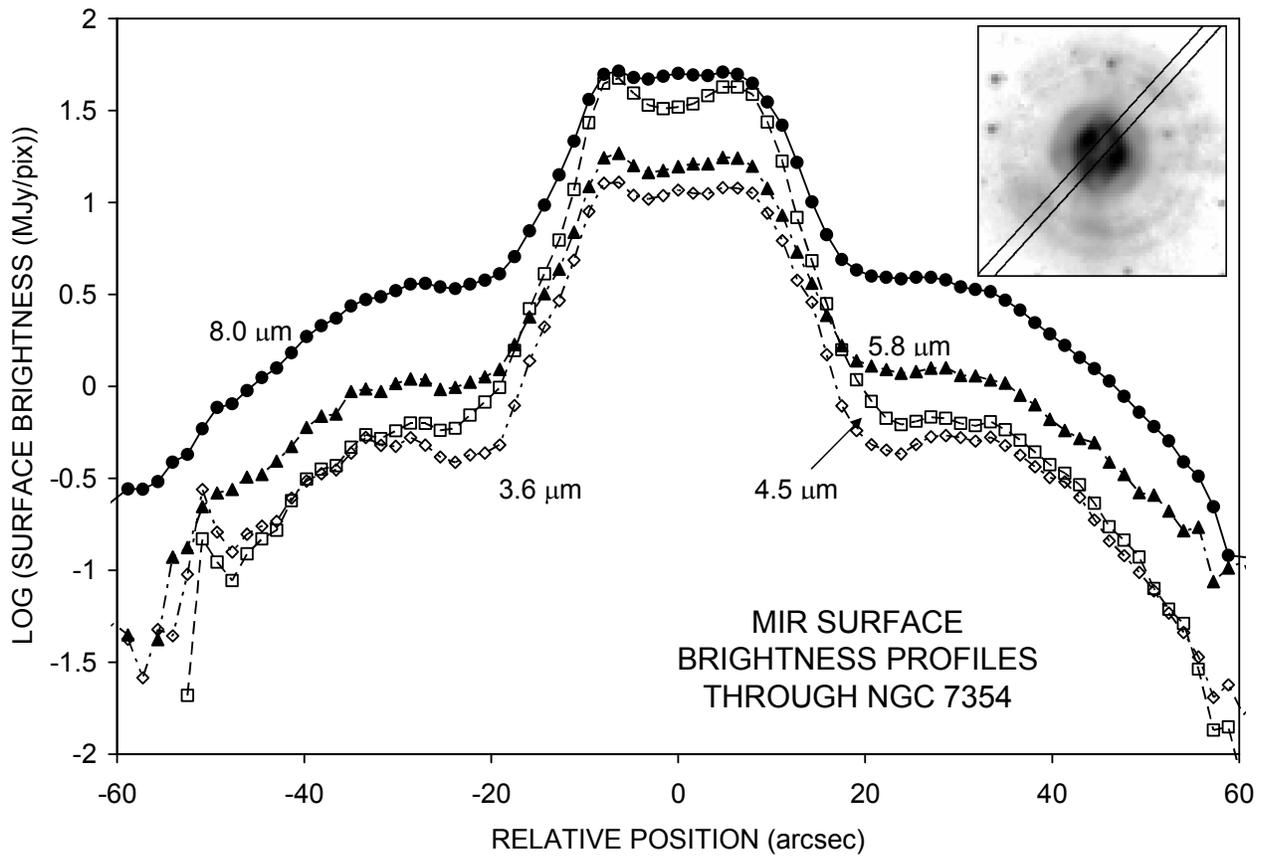
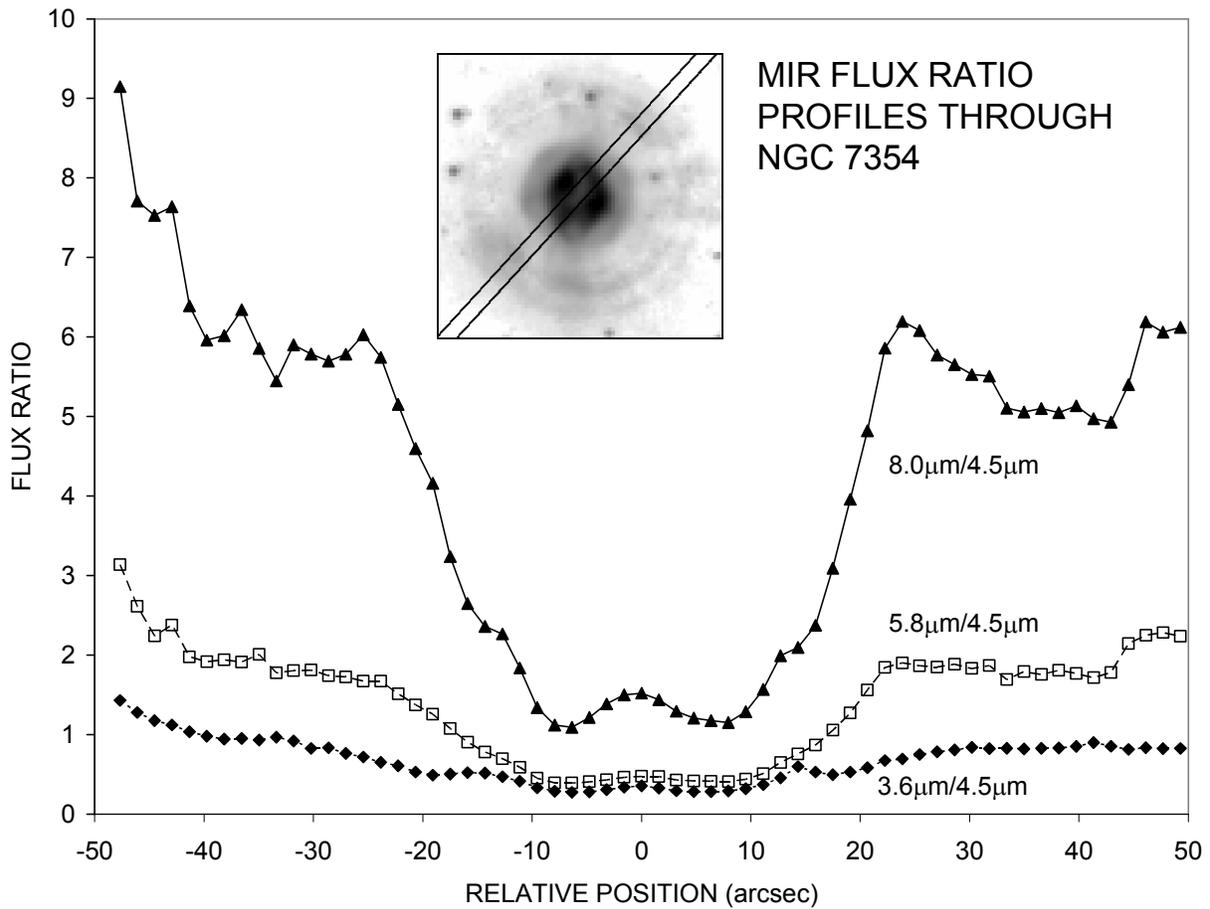

FIGURE 3

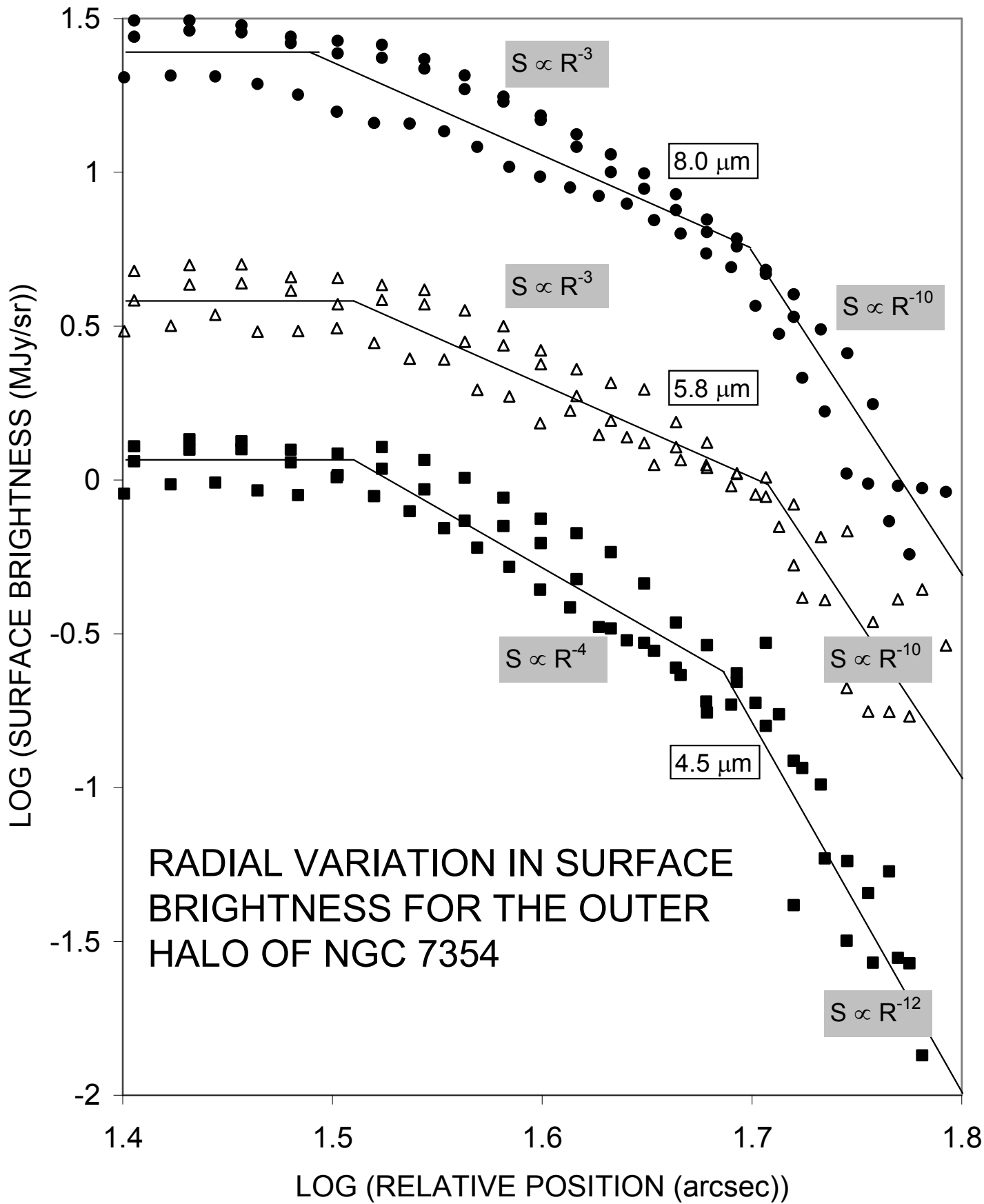

FIGURE 4

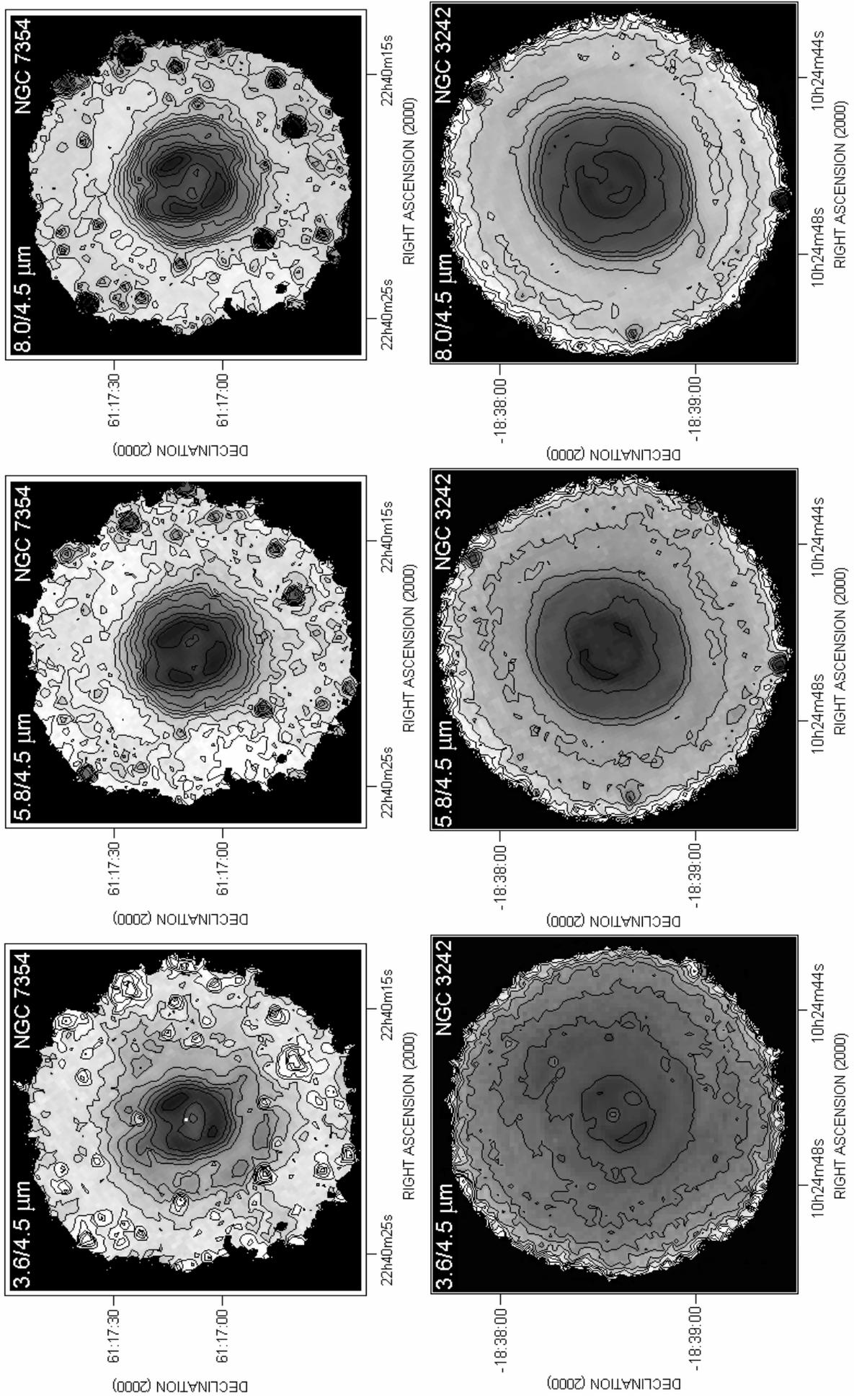

FIGURE 5

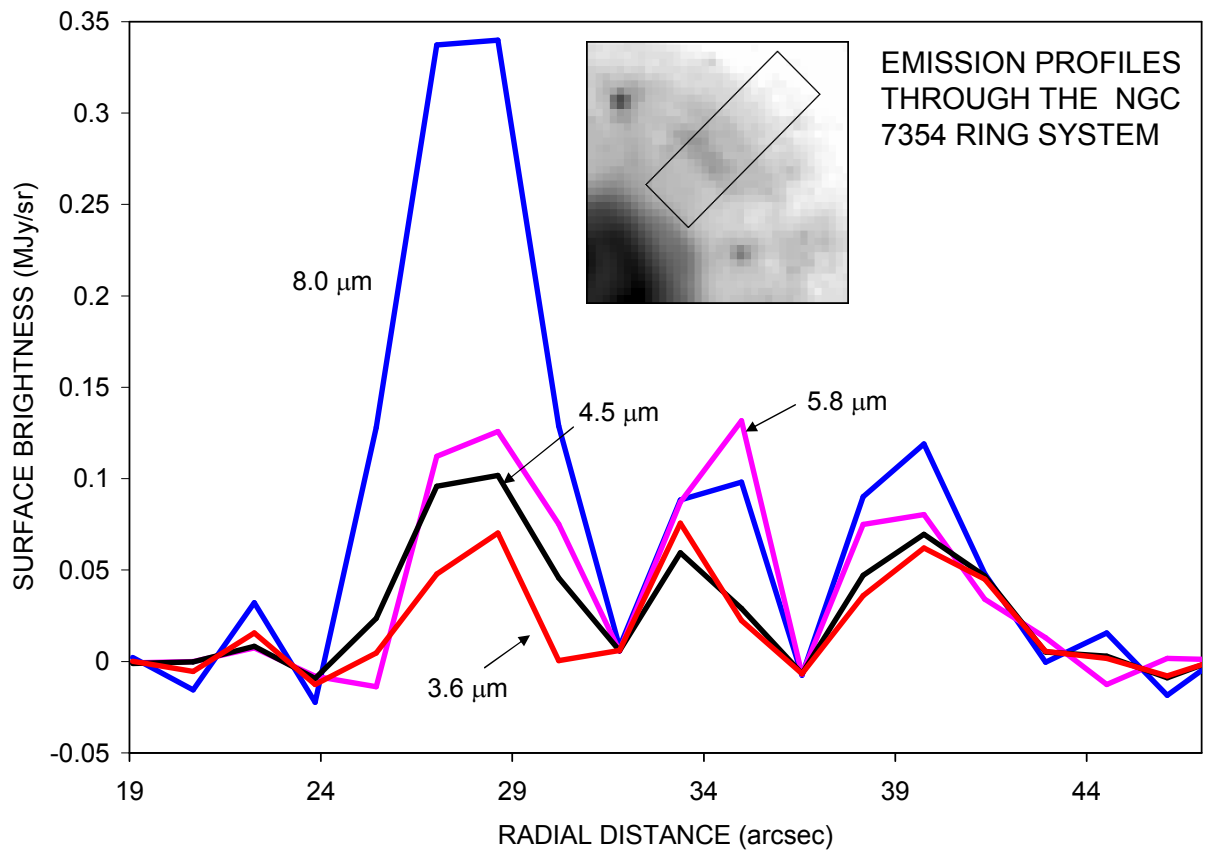
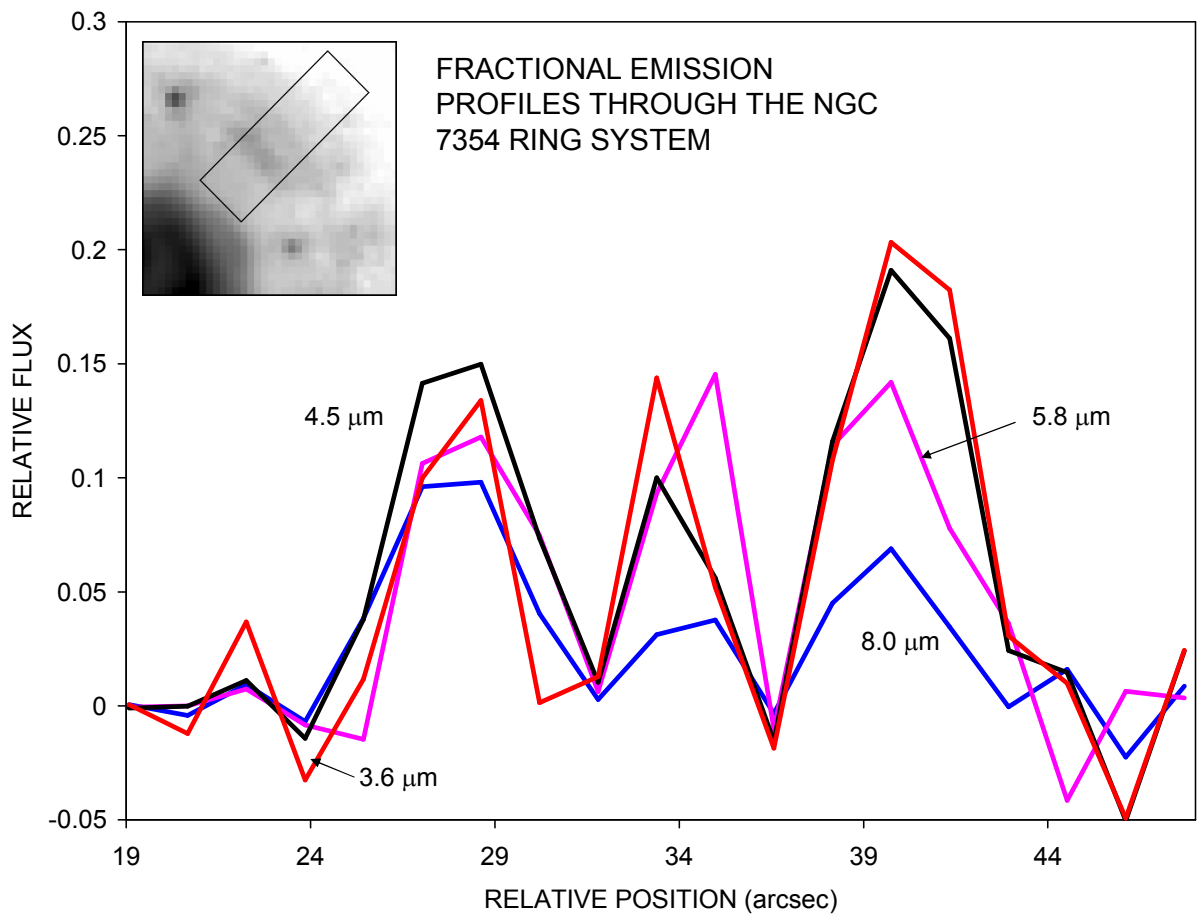

FIGURE 6

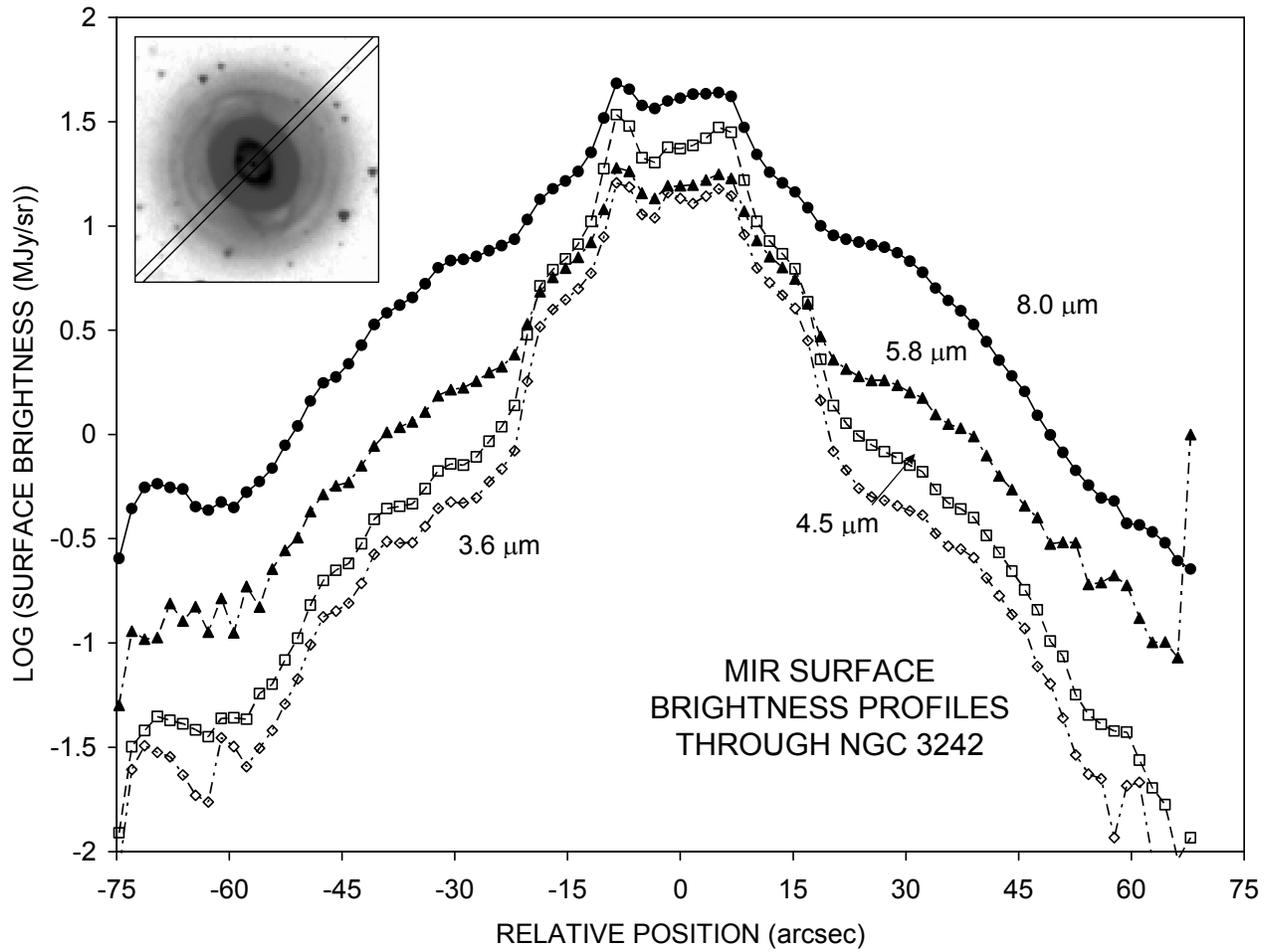
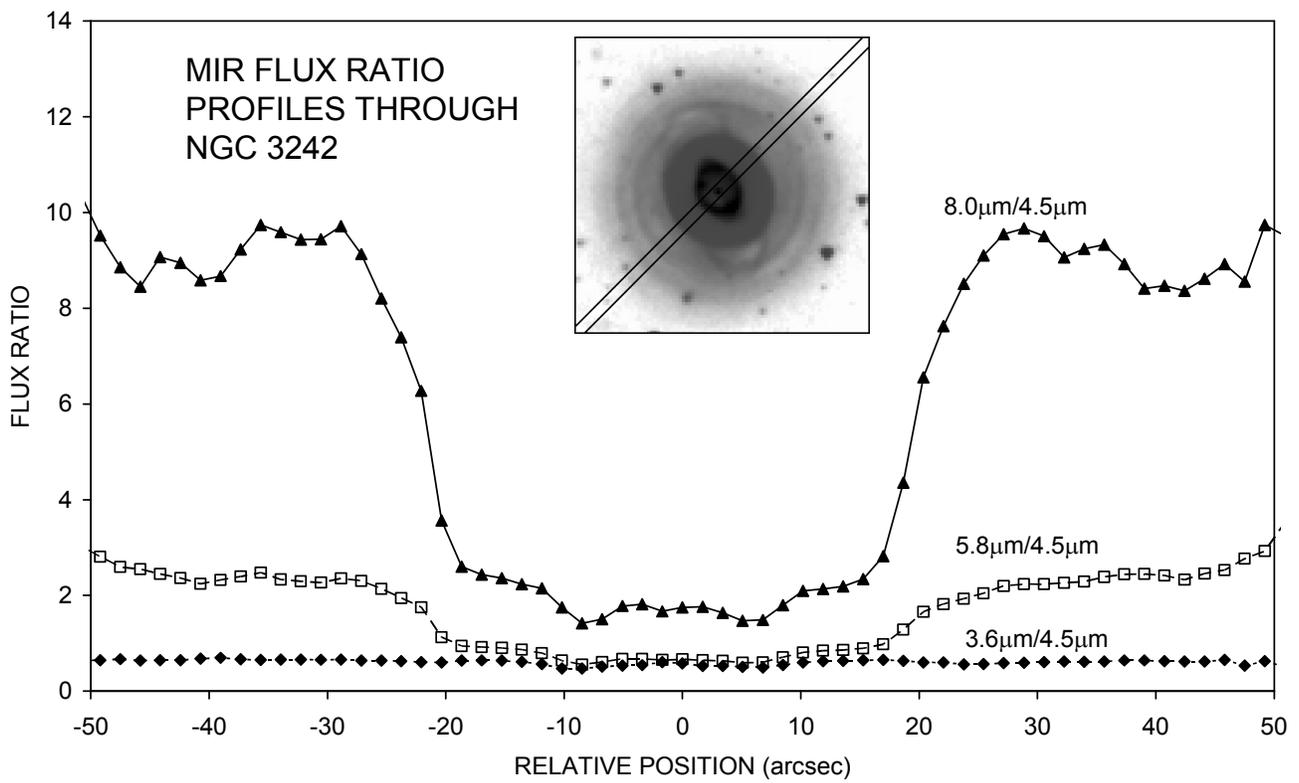

FIGURE 7

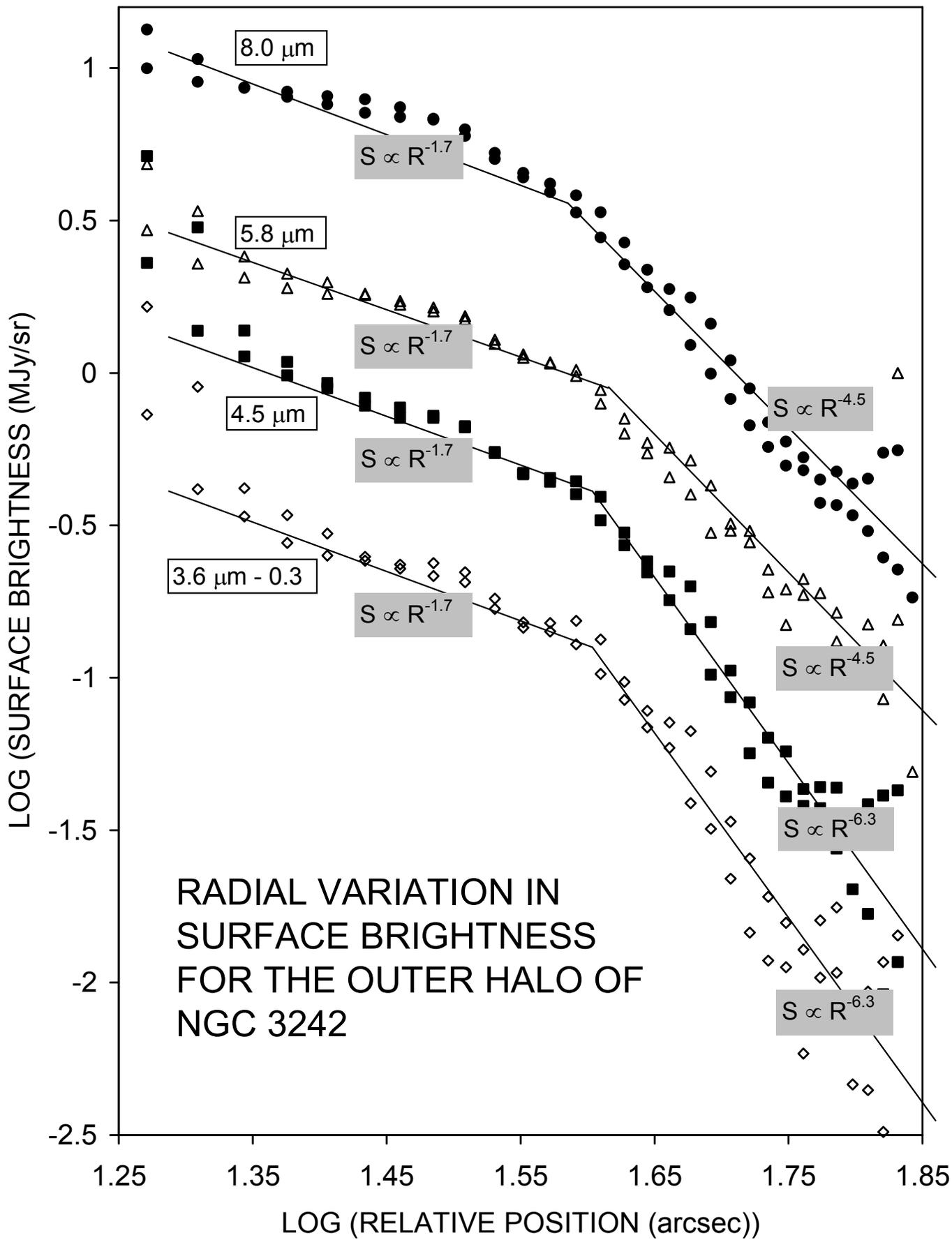

FIGURE 8

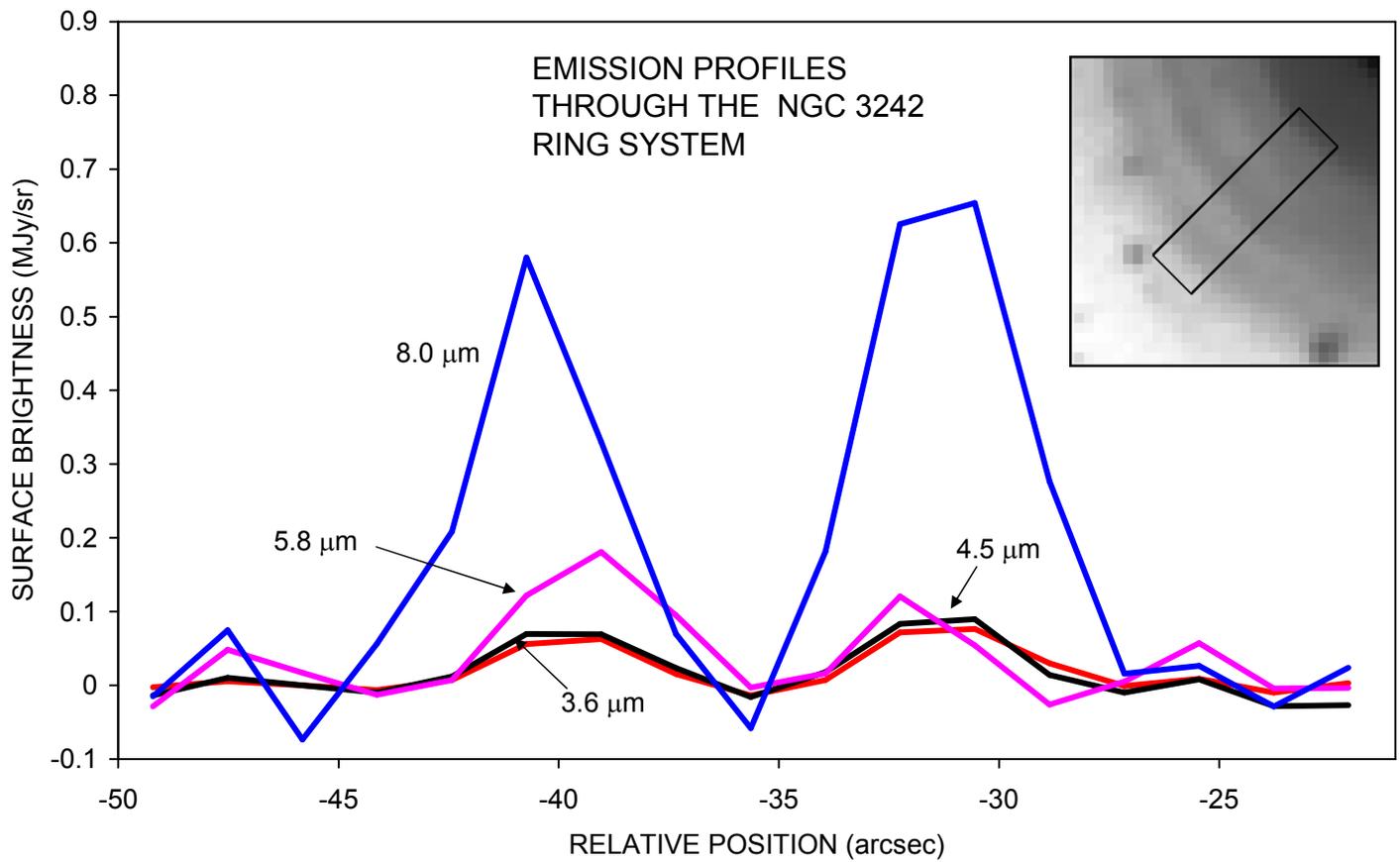
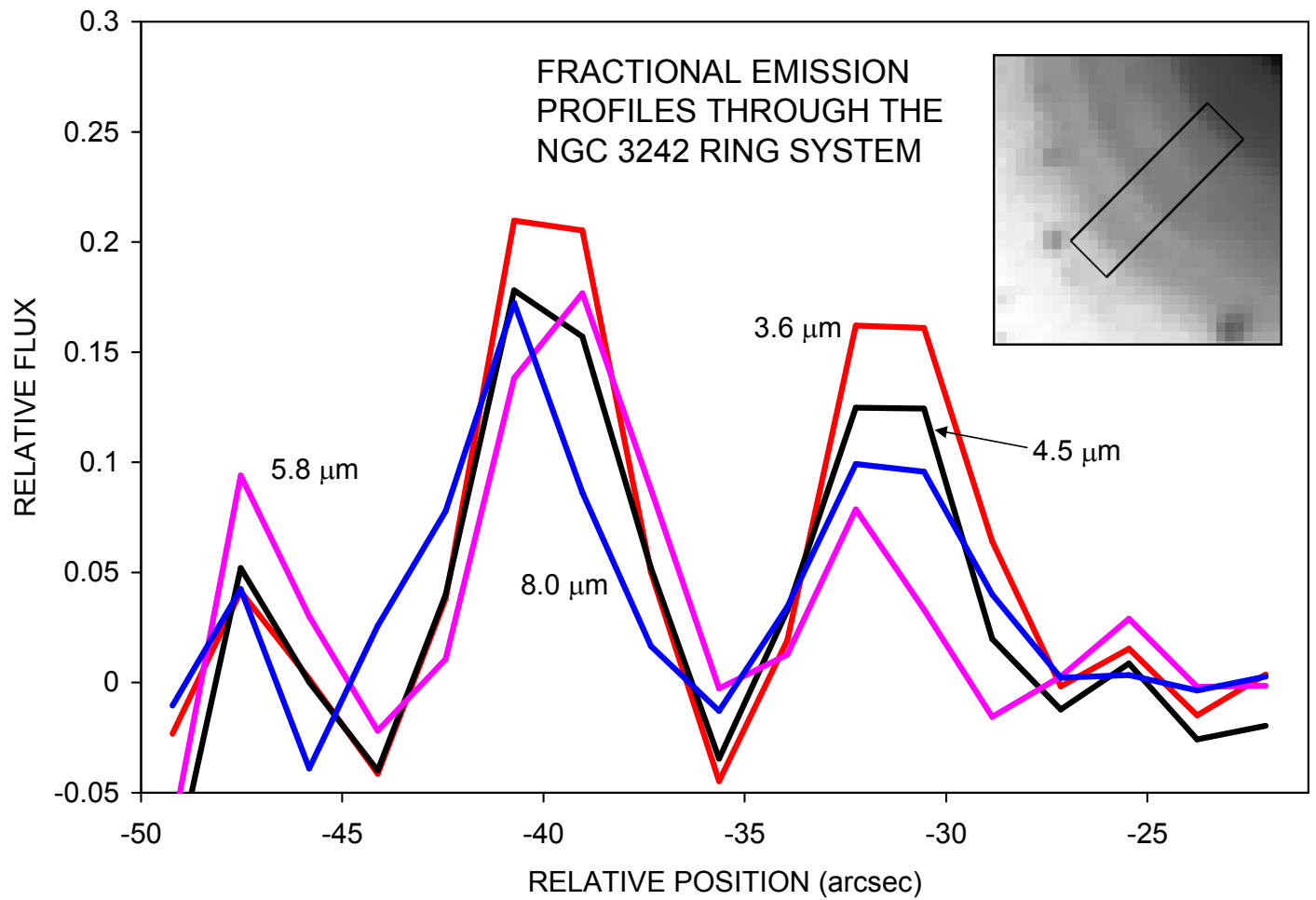

FIGURE 9

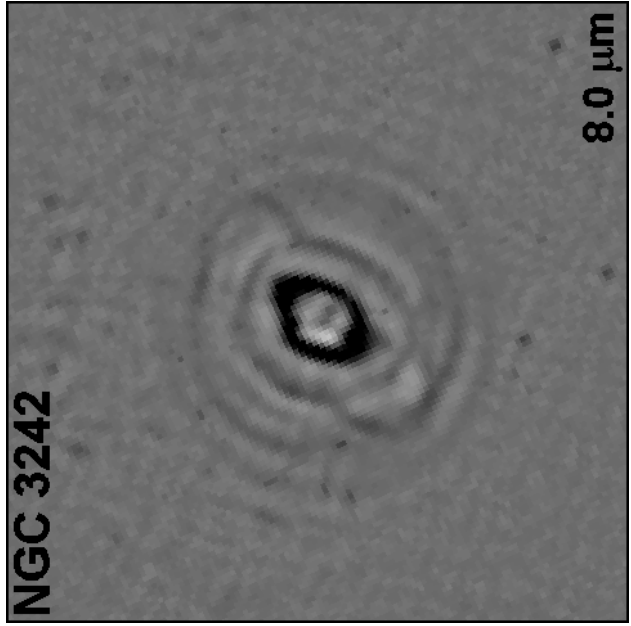
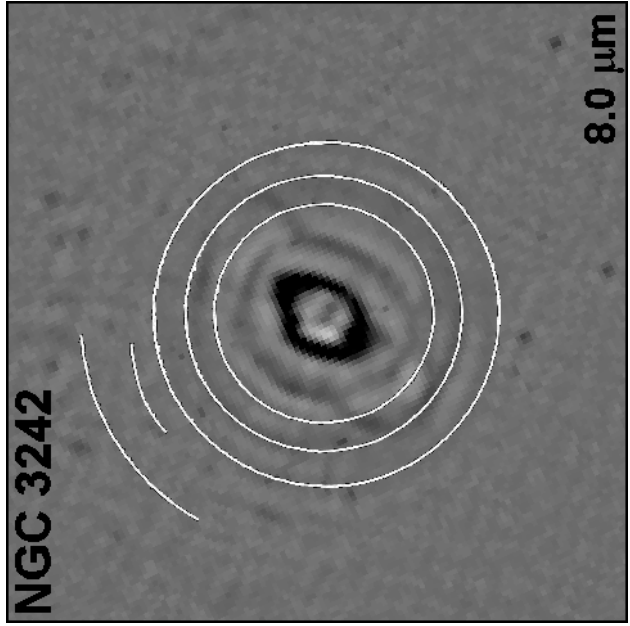
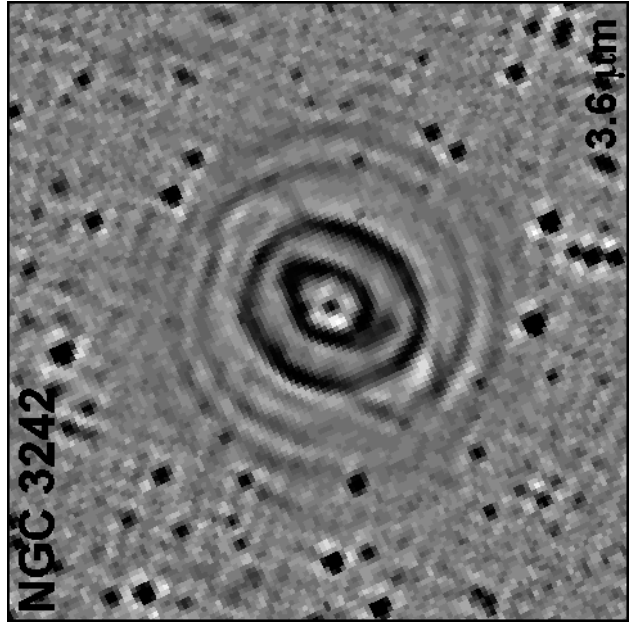
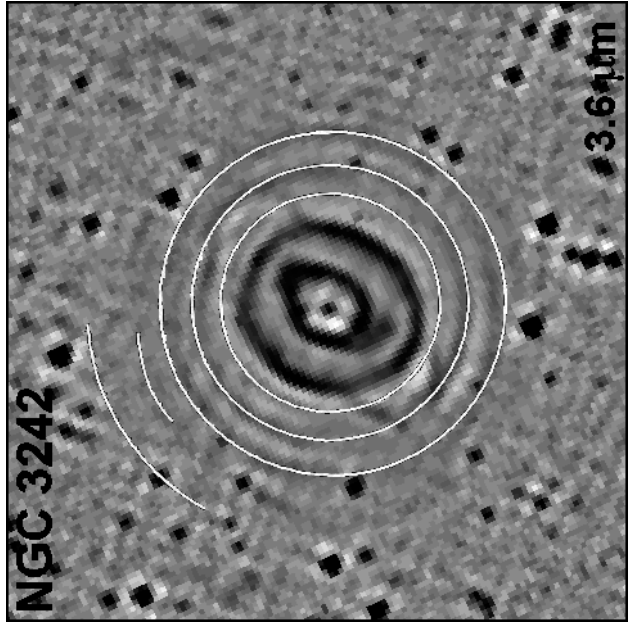
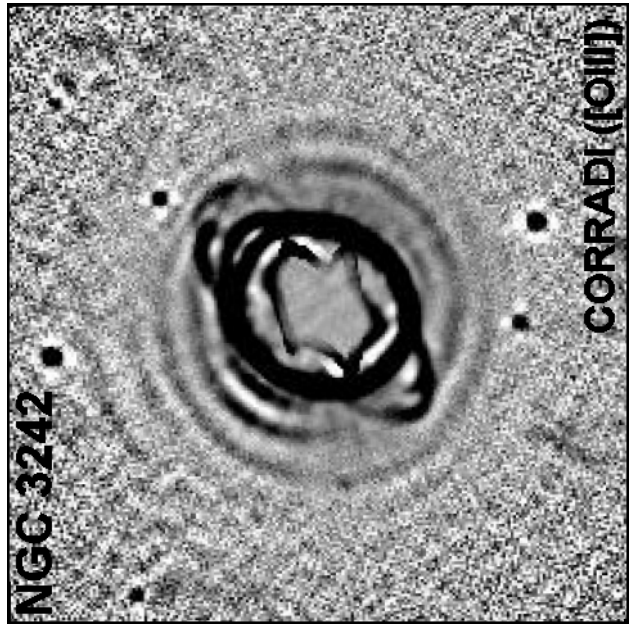
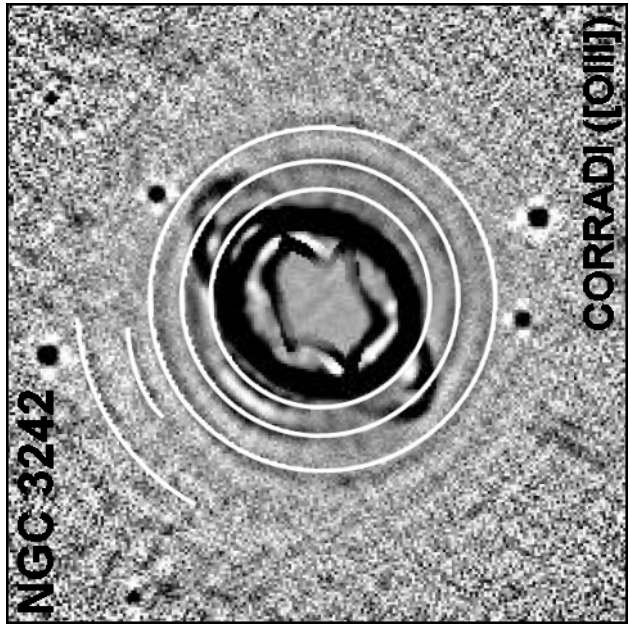

FIGURE 10

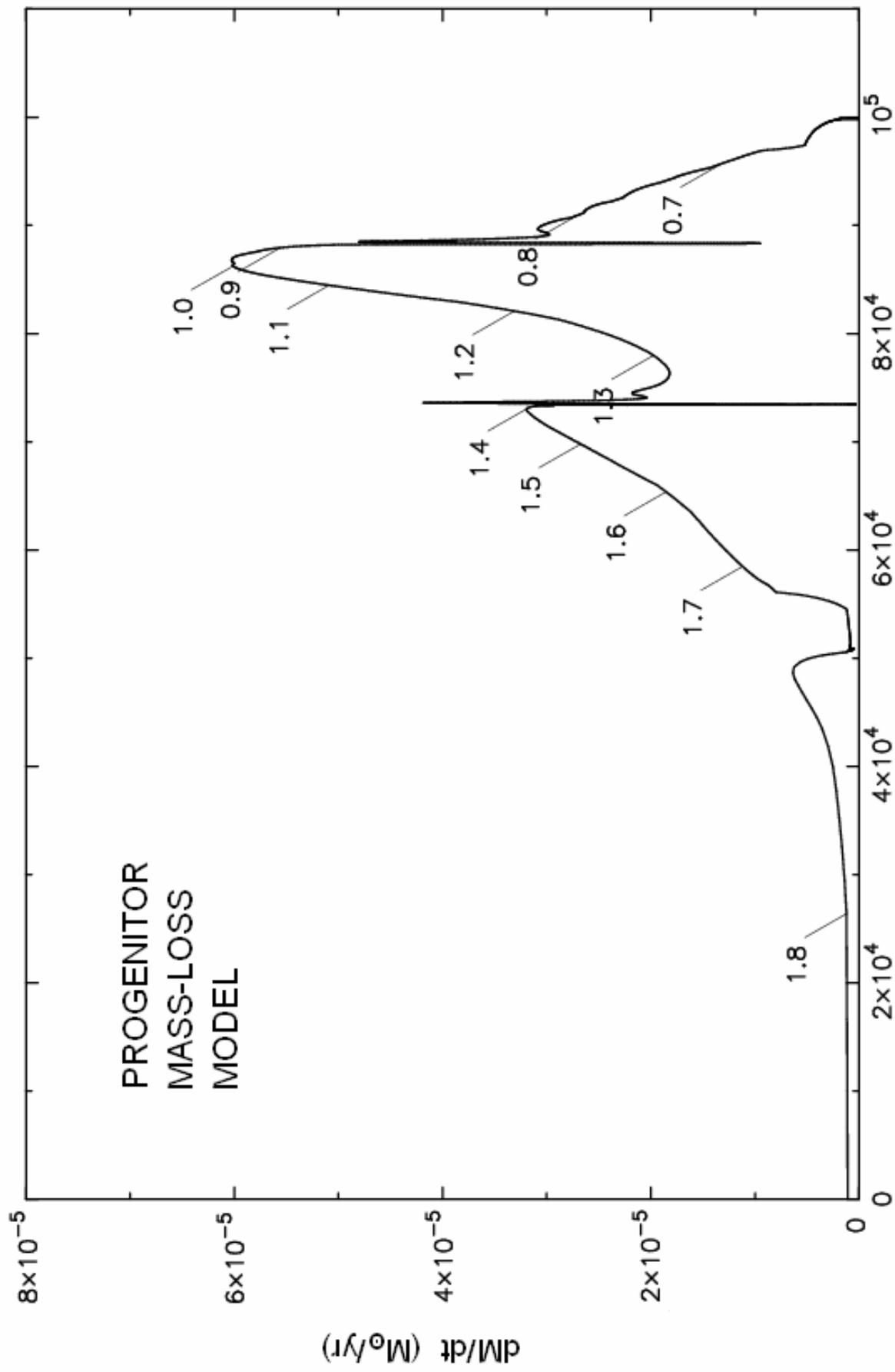

FIGURE 11

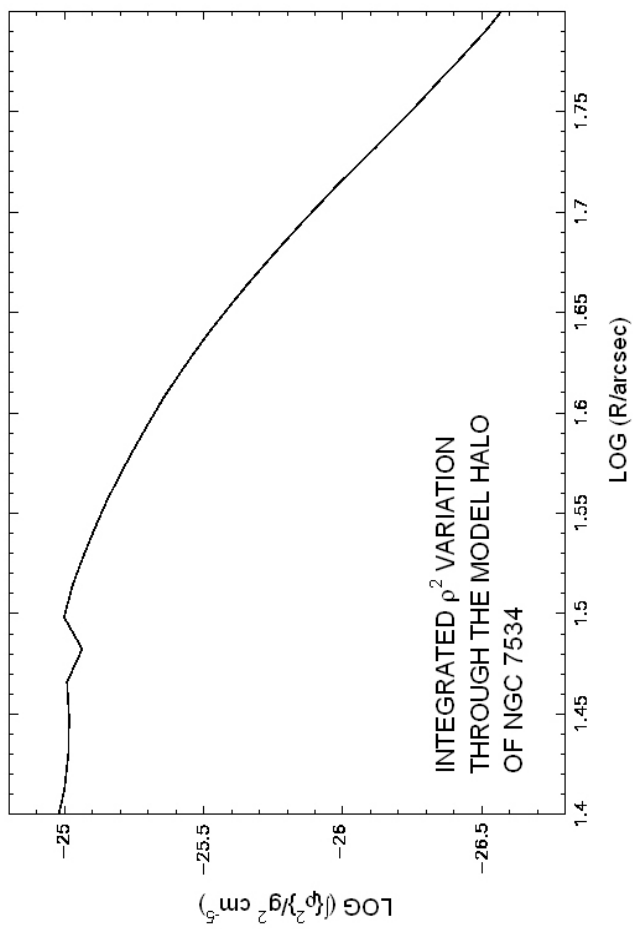
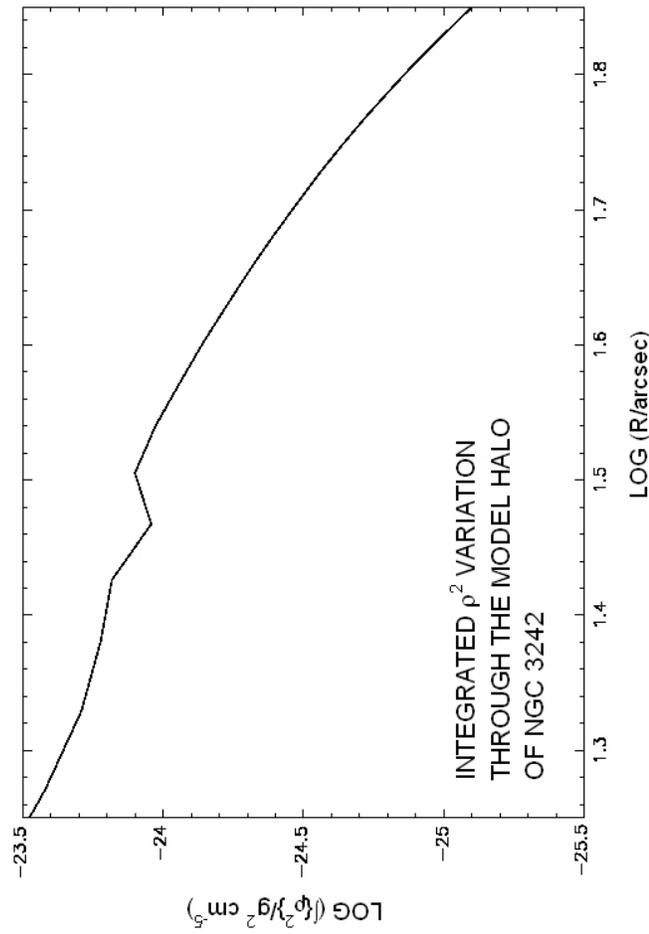
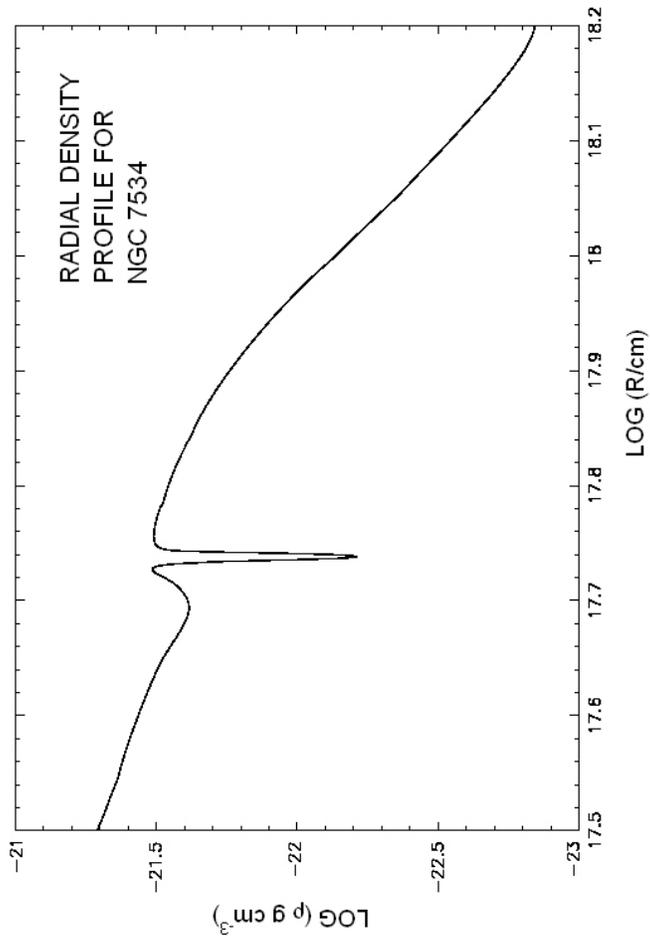
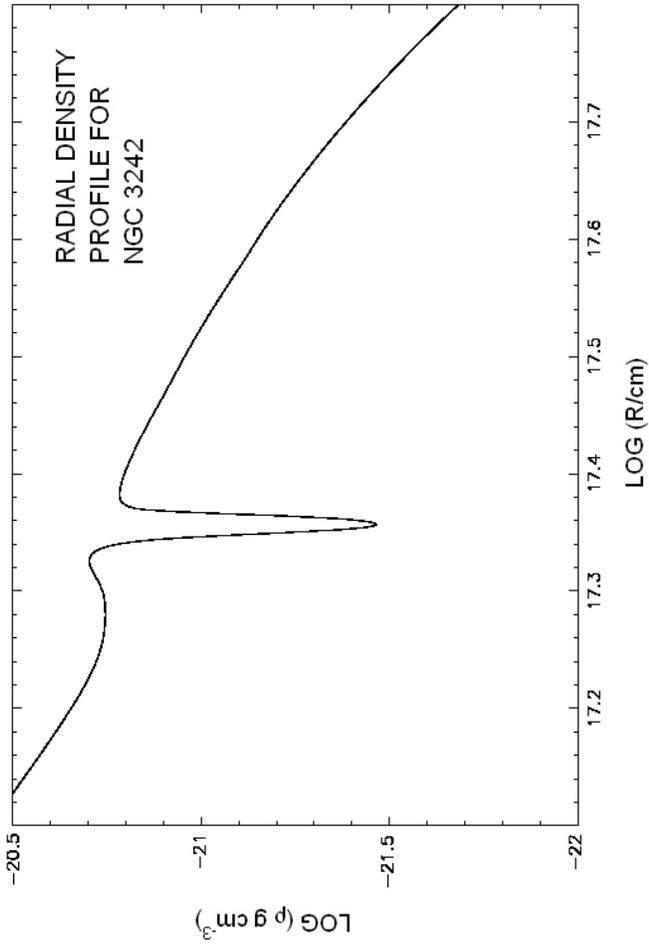

FIGURE 12

56